\documentclass[DIV=12]{scrartcl}
\usepackage{amsmath,amsfonts,amssymb,bm,graphicx,hyperref,verbatim,mathrsfs,tcolorbox,units,enumerate}
\usepackage[numbers,sort&compress]{natbib}
\DeclareOldFontCommand{\rm}{\normalfont\rmfamily}{\mathrm}
\title{Introduction to Quantum Thermodynamics}
\author{Patrick P. Potts}
\date{}

\begin{document}

\maketitle

\begin{abstract}
The theory of quantum thermodynamics investigates how the concepts of heat, work, and temperature can be carried over to the quantum realm, where fluctuations and randomness are fundamentally unavoidable. Of particular practical relevance is the investigation of quantum thermal machines: Machines that use the flow of heat in order to perform some useful task. In this lectures series, we give a brief introduction into how the laws of thermodynamics arise from quantum theory and how thermal machines can be described with Markovian quantum master equations. Recent results are illustrated with examples such as a quantum dot heat engine and a qubit entangler. 
\end{abstract}

\tableofcontents

\section{Introduction}
Quantum thermodynamics investigates heat, work, temperature, as well as related concepts in quantum systems. As these concepts are very general, quantum thermodynamics is of relevance for essentially all scenarios involving open quantum systems. This makes the field extremely broad, including well established topics such as thermoelectricity \cite{pekola:2015}, investigating how an electrical current arises from a temperature gradient, as well as novel approaches such as the resource theory of thermodynamics \cite{lostaglio:2018rev}, an approach that has its origins in entanglement theory. The broad nature of the field implies that the quantum thermodynamics community brings together people with very different backgrounds that have common interests. 

This course is meant to provide a short introduction to this diverse and fast-growing field, introducing basic concepts and illustrating them with simple examples.
After this course, you should...
\begin{enumerate}
	\item ...know how the first and second law of thermodynamics emerge from quantum theory,
	\item ...know what a Markovian master equation is and under what assumptions it can be employed,
	\item ...anticipate what happens when coupling simple systems (such as a qubit or a quantum dot) to one or several thermal reservoirs,
	\item ...be able to calculate observable properties of out-of-equilibrium systems such as heat and charge currents. 
\end{enumerate}
There are several good resources that substantially go beyond the material covered in this course. Very recently, a book on the topic was published that is meant to give a snapshot of the field, providing numerous short reviews on different topics by over 100 contributing authors \cite{thermo:book}. A number of good reviews are provided by Refs.~\cite{kosloff:2013,pekola:2015,vinjanampathy:2016,goold:2016,lostaglio:2018rev,mitchison:2019}. These resources are complemented by the focus issue in Ref.~\cite{anders:2017}.

Since this course aims at introducing an advanced topic in a short amount of time, some concepts will be used without introduction. Notably, the density matrix formalism and second quantization will be used throughout the course. See chapter 2.4 in Ref.~\cite{nielsen:book} and chapter 1 in Ref.~\cite{bruus:book} for a good introduction to these respective topics. In addition, basic knowledge of quantum theory and classical thermodynamics is helpful.

\section{Basic concepts}
In this section, we introduce some basic concepts that are used throughout the course. We set $\hbar=1$ throughout the course.
\subsection{The thermal state}
In the grand-canonical ensemble, the thermal state (Gibbs state) is given by
\begin{equation}
\label{eq:thermalstate}
\hat{\tau}_{\beta,\mu}=\frac{e^{-\beta\left(\hat{H}-\mu\hat{N}\right)}}{Z},\hspace{2cm}Z = {\rm Tr}\left\{e^{-\beta\left(\hat{H}-\mu\hat{N}\right)}\right\},
\end{equation}
where $\hat{H}$ denotes the Hamiltonian of the system, $\hat{N}$ the particle number operator, $\beta=1/(k_{\rm B}T)$ the inverse temperature (with $k_{\rm B}$ being the Boltzmann constant), $\mu$ the chemical potential, and $Z$ is called the partition function.
There are different motivations for the physical relevance of the thermal state.
Consider a small part of a large system, where the large system has a fixed energy and fixed number of particles. It can be shown that the small system is described by the thermal state (under some assumptions) \cite{landaulifshitz}. Therefore, if a system can exchange both energy and particles with an environment, the system is expected to be well described by the thermal state when it is in equilibrium with the environment. The mentioned assumptions will be discussed later in the course, when we discuss equilibration in terms of microscopic equations of motion. 

The thermal state can also be motivated from a principle of maximal entropy \cite{jaynes:1957,jaynes:1957ii}. Consider a scenario where the mean energy $\langle \hat{H}\rangle$ and the mean particle number $\langle \hat{N}\rangle$ are given (physically, they are determined by the temperature and the chemical potential of an environment). In this case, the thermal state maximizes the von Neumann entropy (see box). To see this, consider a state $\hat{\rho}$ with the same mean values as the thermal state. We can then write
\begin{equation}
\begin{aligned}
\label{eq:maxentth}
S_{\rm vN}[\hat{\rho}]&=-{\rm Tr}\left\{\hat{\rho}\ln\hat{\rho}\right\}\leq -{\rm Tr}\left\{\hat{\rho}\ln\hat{\tau}_{\beta,\mu}\right\}\\&=-{\rm Tr}\left\{\hat{\rho}\left[-\beta\left(\hat{H}-\mu\hat{N}\right)-\ln Z\right]\right\}=S_{\rm vN}[\hat{\tau}_{\beta,\mu}],
\end{aligned}
\end{equation}
where we have used the inequality
\begin{equation}
\label{eq:relent}
S[\hat{\rho}||\hat{\sigma}]={\rm Tr}\left\{\hat{\rho}\ln\hat{\rho}-\hat{\rho}\ln\hat{\sigma}\right\}\geq 0.
\end{equation}
Here $S[\hat{\rho}||\hat{\sigma}]$ denotes the quantum relative entropy (see box) and the equality is obtained only for $\hat{\rho}=\hat{\sigma}$.

Finally, the thermal state is the only \textit{completely passive} state \cite{goold:2016}. This means that even if we have many copies of a thermal state, its energy cannot be lowered by any unitary operation, i.e.,
\begin{equation}
\label{eq:passive}
{\rm Tr}\left\{\hat{H}_N\hat{U}\hat{\tau}_{\beta,\mu}^{\otimes N}\hat{U}^\dagger\right\}\geq {\rm Tr}\left\{\hat{H}_N\hat{\tau}_{\beta,\mu}^{\otimes N}\right\},\hspace{1cm}\forall N,\,\hat{U},
\end{equation}
where $\hat{H}_N=\hat{H}^{\otimes N}$ denotes the Hamiltonian corresponding to $N$ copies of the thermal state. The last expression can be interpreted as follows: no work can be extracted from $N$ copies of the thermal state. It can be proven, that the thermal state is the only state that fulfills Eq.~\eqref{eq:passive} \cite{goold:2016}.  

\begin{tcolorbox}
\paragraph{Von Neumann entropy}
\begin{equation}
S_{\rm vN}[\hat{\rho}]=-{\rm Tr}\left\{\hat{\rho}\ln\hat{\rho}\right\}=-\sum_j p_j\ln p_j\geq 0\end{equation}
The von Neumann entropy is equal to the Shannon entropy of the distribution given by the eigenvalues $p_j$ of $\hat{\rho}$. The von Neumann entropy can be understood as the average amount of information that we obtain when observing which eigenstate of $\hat{\rho}$ the system is in. If the state $j$ is observed, the amount of information that is gained can be quantified by $-\ln p_j$. Conversely, the entropy quantifies our lack of knowledge about the system state before observation. This notion is made rigorous by Shannon's noiseless coding theorem which states that storing the outcome of such an observation, using any lossless compression scheme, requires on average at least $-\sum_j p_j\log_2 p_j$ bits. Exchanging the logarithm of base two with the natural logarithm corresponds to quantifying information not in \textit{bits} but in \textit{nats}. The logarithm arises from the fact that information should be additive, i.e., a joint observation of outcomes $l$ and $k$ in two identical systems occurs with probability $p_l p_k$ and results in an information gain $-\ln p_l-\ln p_k$ \cite{shannon:1948,nielsen:book}.
\end{tcolorbox}

\begin{tcolorbox}
	\paragraph{Quantum relative entropy}
	\begin{equation}
	S[\hat{\rho}||\hat{\sigma}]={\rm Tr}\left\{\hat{\rho}\ln\hat{\rho}-\hat{\rho}\ln\hat{\sigma}\right\}=-S_{\rm vN}[\hat{\rho}]-{\rm Tr}\left\{\hat{\rho}\ln\hat{\sigma}\right\}\geq 0\end{equation}
	The quantum relative entropy provides a measure for the information that is lost if a system that is actually in the state $\hat{\rho}$ is erroneously described by the state $\hat{\sigma}$. To see this, consider the classical version of the quantum relative entropy called the Kullback-Leibler divergence $\sum_j p_j\ln p_j/q_j$, where $p_j$ describes the true probability distribution and $q_j$ the one used for describing the system. The information gain we assign to outcome $j$ is given by $-\ln q_j$. The average information gain thus reads $-\sum_j p_j\ln q_j$ and can also be understood as the lack of information about the system before observation. This lack of information includes two parts: the fact that the system was not yet observed, as well as the fact that the wrong distribution is being used to describe it. While the former part is given by the Shannon entropy, the latter part is given by the Kullback-Leibler divergence \cite{kullback:1951,vedral:2002}.
\end{tcolorbox}

\subsection{Non-interacting particles}
Throughout this course, we will mostly consider environments which can be described by non-interacting particles. Either bosons (e.g., electromagnetic-radiation - photons or lattice vibrations - phonons) or fermions (e.g., electrons). 
\subsubsection{Bosons}
A system of non-interacting bosons can be described by the Hamiltonian and number operator
\begin{equation}
\label{eq:hbos}
\hat{H}=\sum_k\varepsilon_k\hat{a}_k^\dagger\hat{a}_k,\hspace{2cm}\hat{N}=\sum_k\hat{a}_k^\dagger\hat{a}_k,
\end{equation}
where $\hat{a}_k$ destroys a particle in state $k$ and fulfills the standard commutation relations
\begin{equation}
\label{eq:boscom}
[\hat{a}_k,\hat{a}_q^\dagger]=\delta_{k,q},\hspace{2cm}[\hat{a}_k,\hat{a}_q]=[\hat{a}_k^{\dagger},\hat{a}_q^{\dagger}]=0,
\end{equation}
where the commutator is defined as $[\hat{A},\hat{B}]=\hat{A}\hat{B}-\hat{B}\hat{A}$.
In a thermal state, Eq.~\eqref{eq:thermalstate}, the mean number of particles in mode $k$ is given by the Bose-Einstein distribution
\begin{equation}
\label{eq:bosedist}
{\rm Tr}\left\{\hat{a}_k^\dagger\hat{a}_k\hat{\tau}_{\beta,\mu}\right\}=n_{\rm B}(\varepsilon_k)=\frac{1}{e^{\beta(\varepsilon_k-\mu)}-1},
\end{equation}
which implies $\varepsilon_k>\mu$ for all $k$ in order to ensure positive occupation numbers.

\subsubsection{Fermions}
A system of non-interacting fermions can be described by the Hamiltonian and number operator
\begin{equation}
\label{eq:hferm}
\hat{H}=\sum_k\varepsilon_k\hat{c}_k^\dagger\hat{c}_k,\hspace{2cm}\hat{N}=\sum_k\hat{c}_k^\dagger\hat{c}_k,
\end{equation}
where $\hat{c}_k$ destroys a particle in state $k$ and fulfills the standard anti-commutation relations
\begin{equation}
\label{eq:fercom}
\{\hat{c}_k,\hat{c}_q^\dagger\}=\delta_{k,q},\hspace{2cm}\{\hat{c}_k,\hat{c}_q\}=\{\hat{c}_k^{\dagger},\hat{c}_q^{\dagger}\}=0,
\end{equation}
where the anti-commutator is defined as $\{\hat{A},\hat{B}\}=\hat{A}\hat{B}+\hat{B}\hat{A}$.
In a thermal state, Eq.~\eqref{eq:thermalstate}, the mean number of particles in mode $k$ is given by the Fermi-Dirac distribution
\begin{equation}
\label{eq:fermidist}
{\rm Tr}\left\{\hat{c}_k^\dagger\hat{c}_k\hat{\tau}_{\beta,\mu}\right\}=n_{\rm F}(\varepsilon_k)=\frac{1}{e^{\beta(\varepsilon_k-\mu)}+1}.
\end{equation}
For fermions, the single particle energies can lie both above as well as below the chemical potential.

{\small
\paragraph{Exercises}

\paragraph{1.1 Bose-Einstein and Fermi-Dirac distributions (2pt)}

Show that, in the absence of interactions, the mean occupation of a given mode in thermal equilibrium is determined by the Bose-Einstein distribution for bosons and by the Fermi-Dirac distribution for fermions (i.e., show Eqs.~\eqref{eq:bosedist} and \eqref{eq:fermidist} when the state of the system is given by Eq.~\eqref{eq:thermalstate} with the Hamiltonian and number operator given by Eq.~\eqref{eq:hbos} and Eq.~\eqref{eq:hferm} respectively).

\paragraph{1.2 Thermal state of a qubit and a quantum dot (2pt)}

Explicitly calculate the thermal state of a qubit (two-level system) at temperature $T$ with Hamiltonian
\begin{equation}
\label{eq:hamqubit}
\hat{H}_{\rm q} =  \frac{\varepsilon_{\rm q}}{2} \hat{\sigma}_{\rm z} = \begin{pmatrix}
\varepsilon_{\rm q}/2 & 0\\
0 & -\varepsilon_{\rm q}/2
\end{pmatrix}.
\end{equation}
Compare to the thermal state of a quantum dot (single level fermionic system) at temperature $T$ and chemical potential $\mu$ with Hamiltonian
\begin{equation}
\label{eq:hamdot}
\hat{H}_{\rm d} = \varepsilon_{\rm d} \hat{c}^\dagger\hat{c}.
\end{equation}

\paragraph{1.3 Thermal state of a spinful quantum dot (2pt)}

Explicitly calculate the thermal state of a spinful quantum dot at temperature $T$ and chemical potential $\mu$ with Hamiltonian
\begin{equation}
\label{eq:hamdotspin}
\hat{H}_{\rm sd} = \varepsilon_{\rm d} (\hat{c}_\uparrow^\dagger\hat{c}_\uparrow+\hat{c}_\downarrow^\dagger\hat{c}_\downarrow)+U\hat{c}_\uparrow^\dagger\hat{c}_\uparrow\hat{c}_\downarrow^\dagger\hat{c}_\downarrow.
\end{equation}
What happens in the limits $U\rightarrow 0$ and $U\rightarrow\infty$?}

\paragraph{1.4 Non-negativity of the quantum relative entropy (2pt)}

Show the inequality
\begin{equation}
S[\hat{\rho}||\hat{\sigma}]\geq 0.
\end{equation}

\paragraph{Hint:} Use Jensen's inequality. A function $f$ is called convex if
\begin{equation}
\label{eq:convex}
f(tx_1+(1-t)x_2)\leq tf(x_1)+(1-t)f(x_2),
\end{equation}
for all $x_1, x_2$ and $t\in[0,1]$. Jensen's inequality states that if $x$ denotes a random variable and $f$ a convex function, then
\begin{equation}
f(E[x])\leq E[f(x)],
\end{equation}
where $E[\cdot]$ denotes the expectation value. For a discrete probability distribution $p_j\geq0$ and $\sum_j p_j=1$, the inequality reads [note the similarity to Eq.~\eqref{eq:convex}]
\begin{equation}
\label{eq:exp}
f\left(\sum_j p_jx_j\right)\leq  \sum_j p_j f(x_j).
\end{equation}

\section{The laws of thermodynamics}
\label{sec:laws}

In this section, we discuss how the laws of thermodynamics arise in quantum mechanics.

\subsection{The general scenario}
\label{sec:general}
The general scenario we consider consists of a system coupled to multiple baths which are in (or close to) thermal equilibrium. This is described by the Hamiltonian
\begin{equation}
\label{eq:htot}
\hat{H}_{\rm tot}(t)=\hat{H}_{\rm S}(t)+\sum_{\alpha}\left(\hat{H}_\alpha+\hat{H}_{\alpha \rm S}\right),
\end{equation}
where the system is labeled by the subscript S and the baths are labeled by the index $\alpha$. the term $\hat{H}_{\alpha \rm S}$ denotes the coupling between the system and bath $\alpha$. The time evolution of the total density matrix is then given by
\begin{equation}
\label{eq:rhototu}
\hat{\rho}_{\rm tot}(t)=\hat{U}(t)\hat{\rho}_{\rm tot}(0)\hat{U}^\dagger(t)\hspace{1cm}\Leftrightarrow\hspace{1cm}\partial_t\hat{\rho}_{\rm tot}(t)=-i[\hat{H}_{\rm tot}(t),\hat{\rho}_{\rm tot}(t)],
\end{equation}
where the time-evolution operator is given by
\begin{equation}
\label{eq:timeev}
\hat{U}(t)=\mathcal{T}e^{-i\int_0^t dt'\hat{H}_{\rm tot}(t')},
\end{equation}
with $\mathcal{T}$ denoting the time-ordering operator (see box) and we allow the system Hamiltonian to be time-dependent.
\begin{tcolorbox}
	\paragraph{Time ordered exponential}
The time ordered exponential in Eq.~\eqref{eq:timeev} can be written as
\begin{equation}
\begin{aligned}
\mathcal{T}e^{-i\int_0^t dt'\hat{H}_{\rm tot}(t')} = \lim_{\delta t\rightarrow 0}e^{-i\delta t\hat{H}_{\rm tot}(t)}&e^{-i\delta t\hat{H}_{\rm tot}(t-\delta t)}e^{-i\delta t\hat{H}_{\rm tot}(t-2\delta t)}\cdots\\&\cdots e^{-i\delta t\hat{H}_{\rm tot}(2\delta t)}e^{-i\delta t\hat{H}_{\rm tot}(\delta t)}e^{-i\delta t\hat{H}_{\rm tot}(0)},
\end{aligned}
\end{equation}
such that the time argument in the Hamiltonian on the right-hand side increases from the right to the left of the expression. Each exponential in the product can be understood as the time-evolution by the infinitesimal time-step $\delta t$ \cite{puri:book}.
\end{tcolorbox}

Throughout this course, we will mostly be interested in how energy flows through the system between the different reservoirs. To this end, it is instructive to consider the mean energy change of reservoir $\alpha$
\begin{equation}
\label{eq:enalp}
\partial_t\langle \hat{H}_\alpha\rangle={\rm Tr}\left\{\hat{H}_\alpha\partial_t\hat{\rho}_{\rm tot}\right\}=\partial_t\langle \hat{H}_\alpha-\mu_\alpha\hat{N}_\alpha\rangle+\mu_\alpha\partial_t\langle \hat{N}_\alpha\rangle=J_\alpha+P_\alpha.
\end{equation}
Here we divided the energy change into a part that we call the heat current $J_\alpha$ and a part that we call the power $P_\alpha$ that enters reservoir $\alpha$. To motivate this separation of energy flows into heat and work, we need to introduce the concept of entropy.

\subsection{Entropy}
\label{sec:entropy}
The standard extension of the entropy to the quantum regime is the von Neumann entropy.
However, under unitary time-evolution, the von Neumann entropy of the total density matrix is invariant as a function of time
\begin{equation}
\label{eq:vnSt}
\partial_t S_{\rm vN}[\hat{\rho}_{\rm tot}]=-\partial_t{\rm Tr}\{\hat{\rho}_{\rm tot}(t)\ln \hat{\rho}_{\rm tot}(t)\}=0.
\end{equation}
The von Neumann entropy of the total system can thus not tell us anything about how energy flows between systems and reservoirs. To make progress, we consider an effective description based on local thermal equilibrium [i.e., the reservoirs are described by the thermal states given in Eq.~\eqref{eq:thermalstate}]
\begin{itemize}
	\item True description: $\hat{\rho}_{\rm tot}(t)$
	\item Effective description: $\hat{\rho}_{\rm S}(t)\bigotimes_\alpha\hat{\tau}_{\beta_\alpha,\mu_\alpha}$
\end{itemize}
with $\hat{\rho}_{\rm S}={\rm Tr}_{\rm \bar{S}}\{\hat{\rho}_{\rm tot}\}$ where the trace is taken over all but the system's degrees of freedom. The effective description thus contains all information on the system but neglects any changes to the reservoirs as well as the correlations between system and reservoirs. Such an effective description is often the best one can do in an experiment, where one might have control over the microscopic degrees of freedom of the system only.

We now consider the information that is lost when using the effective description instead of the true description. The information that is lost when describing a system by an inaccurate density matrix is given by the quantum relative entropy (see box)
\begin{equation}
\label{eq:qreeff}
S\left[{\textstyle \hat{\rho}_{\rm tot}||\hat{\rho}_{\rm S}\bigotimes_\alpha\hat{\tau}_{\beta_\alpha,\mu_\alpha}}\right]=S_{\rm vN}[\hat{\rho}_{\rm S}]+\sum_{\alpha}\beta_\alpha \langle \hat{H}_\alpha-\mu_\alpha\hat{N}_\alpha\rangle+\sum_{\alpha}\ln Z_{\alpha}-S_{\rm vN}[\hat{\rho}_{\rm tot}].
\end{equation}
In particular, we find
\begin{equation}
\label{eq:entprod}
\Sigma = \partial_t S\left[{\textstyle \hat{\rho}_{\rm tot}||\hat{\rho}_{\rm S}\bigotimes_\alpha\hat{\tau}_{\beta_\alpha,\mu_\alpha}}\right]=\partial_tS_{\rm vN}[\hat{\rho}_{\rm S}]+\frac{1}{k_{\rm B}}\sum_{\alpha}\frac{J_\alpha}{T_\alpha}.
\end{equation}
This expression will be used as the \textit{entropy production rate} throughout the lecture \cite{esposito:2010njp}. It can be interpreted as the amount of information lost by our local equilibrium description, due to the build up of correlations between system and bath as well as changes to the reservoirs. Note that it is not guaranteed to be positive. Finite size effects as well as non-Markovian dynamics can result in a negative entropy production (a backflow of information from the bath). However, for infinitely large and memoryless baths, the entropy production is ensured to be positive at all times as the information is irretrievably lost when one can only access the system alone (more on this later).

Note that Eq.~\eqref{eq:entprod} motivates the interpretation of $J_\alpha$ as a heat flow, such that the entropy production associated to bath $\alpha$ is given by the usual expression for baths which remain in thermal equilibrium. To see this, consider a bath which exchanges energy and/or particles with a system but nevertheless remains in a thermal state at all times. Physically, we assume that whenever a particle enters the bath, the bath immediately re-thermalizes with a slightly higher value for $\langle \hat{N}\rangle$ (and similarly for energy). The temperature and chemical potential may thus vary in time.
%In this scenario, we find
%\begin{equation}
%\label{eq:bathent}
%\partial_tS_{\rm vN}[\hat{\tau}_{\beta_t,\mu_t}]=\partial_t\beta_t\langle \hat{H}-\mu_t\hat{N}\rangle+\partial_t\ln Z =\beta_t\partial_t\langle \hat{H}\rangle-\beta_t\mu_t\partial_t\langle \hat{N}\rangle,
%\end{equation}
%where we used
%\begin{equation}
%\partial_t\ln Z =-\langle \hat{H}\rangle\partial_t\beta_t+\langle\hat{N}\rangle\partial_t(\beta_t\mu_t).
%\end{equation}
If the variations of $\beta_t$ and $\mu_t$ are determined by small deviations from the values $\beta$ and $\mu$, it can be shown that (see Ex.\,2.1)
\begin{equation}
\label{eq:bathent2}
\partial_tS_{\rm vN}[\hat{\tau}_{\beta_t,\mu_t}]=\beta\partial_t\langle \hat{H}-\mu\hat{N}\rangle=\frac{J}{k_{\rm B}T}.
\end{equation}

\subsection{First law of thermodynamics}
\label{sec:firstlaw}
To write down the first law of thermodynamics in terms of the particle and energy change in the system, we make a few simplifying assumptions. First, we consider particle number conservation
\begin{equation}
\label{eq:partcons}
\left[\hat{N}_{\rm S}+\sum_{\alpha}\hat{N}_\alpha,\hat{H}_{\rm tot}\right]=0,\hspace{1cm}\left[\hat{N}_\alpha,\hat{H}_{\alpha}\right]=0,\hspace{1cm}\left[\hat{N}_{\rm S},\hat{H}_{\rm S}\right]=0,
\end{equation}
where the first equation ensures global and the following equations ensure local particle conservation. From these equations, we find
\begin{equation}
\label{eq:partsa}
[\hat{N}_\alpha+\hat{N}_{\rm S},\hat{H}_{\alpha \rm S}]=0,
\end{equation}
which implies that particles that leave bath $\alpha$ have to end up in the system.

We will further assume that the energy stored in the coupling does not change over time
\begin{equation}
\label{eq:intsa}
\partial_t\langle \hat{H}_{\alpha \rm S}\rangle=-i\langle [\hat{H}_{\alpha \rm S},\hat{H}_\alpha+\hat{H}_{\rm S}]\rangle=0.
\end{equation}
This is obviously true if the commutator in the last expression vanishes (i.e., for so-called resonant interactions). It is also true in the small coupling limit, where any changes in the coupling energy can be neglected in comparison with the energy changes in the system and baths.

With Eqs.~\eqref{eq:partsa} and \eqref{eq:intsa}, we can write
\begin{equation}
\label{eq:heatworksys}
J_\alpha = i\langle[\hat{H}_{\rm S}-\mu_\alpha \hat{N}_{\rm S},\hat{H}_{\alpha \rm S}]\rangle,\hspace{1.5cm}P_\alpha = i\mu_\alpha \langle[\hat{N}_{\rm S},\hat{H}_{\alpha \rm S}]\rangle,
\end{equation}
relating the power and the heat flow into reservoir $\alpha$ to the change in energy and particle number of the system, mediated by the coupling with bath $\alpha$. The total energy change of the system can then be written as
\begin{equation}
\label{eq:firstlaw}
\partial_t\langle \hat{H}_{\rm S}\rangle = P_{\rm S} -\sum_{\alpha}(J_\alpha+P_\alpha),\hspace{1.5cm}P_{\rm S}=\langle \partial_t\hat{H}_{\rm S}\rangle,
\end{equation}
where $P_{\rm S}$ denotes the power entering the system due to some external classical drive that renders $\hat{H}_{\rm S}$ time-dependent. The last equation is the first law of thermodynamics. Note that the energy flows are defined to be positive when they enter the location corresponding to their index.

\subsection{Second law of thermodynamics}
\label{sec:secondlaw}
Let us consider an initial state which is a product state of the form
\begin{equation}
\label{eq:initst}
\hat{\rho}_{\rm tot}(0)=\hat{\rho}_{\rm S}(0)\bigotimes_\alpha\hat{\tau}_{\beta_\alpha,\mu_\alpha}.
\end{equation}
In this case, Eq.~\eqref{eq:qreeff} can be written as
\begin{equation}
\label{eq:entropy}
\Delta S(t) \equiv S\left[{\textstyle \hat{\rho}_{\rm tot}(t)||\hat{\rho}_{\rm S}(t)\bigotimes_\alpha\tau_{\beta_\alpha,\mu_\alpha}}\right]= S_{\rm vN}[\hat{\rho}_{\rm S}(t)]-S_{\rm vN}[\hat{\rho}_{\rm S}(0)]+\frac{1}{k_{\rm B}}\sum_{\alpha}\frac{Q_\alpha}{T_\alpha},
\end{equation}
where the heat is defined as
\begin{equation}
\label{eq:heat}
Q_\alpha = {\rm Tr} \left\{(\hat{H}_\alpha-\mu_\alpha\hat{N}_\alpha)\hat{\rho}_{\rm tot}(t)\right\}-{\rm Tr} \left\{(\hat{H}_\alpha-\mu_\alpha\hat{N}_\alpha)\hat{\rho}_{\rm tot}(0)\right\}.
\end{equation}
Since it is expressed as a relative quantum entropy, we have
\begin{equation}
\label{eq:secondlaw1}
\Delta S(t)\geq 0.
\end{equation}
From an information point of view, this inequality tells us that if our effective description is true at $t=0$, then it can only be worse at later times.
As mentioned above, the entropy production rate $\partial_t\Delta S(t)$ is not always guaranteed to be positive (i.e., $\Delta S(t)$ is not necessarily a monotonously increasing function of time). However, at small times, Eq.~\eqref{eq:secondlaw1} ensures that the entropy production rate is also positive. 

The above description relies on Eq.~\eqref{eq:initst}, i.e., on the fact that at some time that we call $t=0$, the system and the reservoirs are uncorrelated and the reservoirs truly are in a thermal state. Let us know consider the family of states
\begin{equation}
\label{eq:initst2}
\hat{\rho}^{t_0}_{\rm tot}(t_0)=\hat{\rho}_{\rm S}(t_0)\bigotimes_\alpha\hat{\tau}_{\beta_\alpha,\mu_\alpha}.
\end{equation}
We can choose each of these states as the initial condition for the time-evolution of the total density matrix. However, the reduced state of the system might depend on the chosen $t_0$
\begin{equation}
\label{eq:redt0dep}
\hat{\rho}_{\rm S}^{t_0}(t)={\rm Tr}_{\bar{\rm S}}\{\hat{\rho}^{t_0}_{\rm tot}(t)\}={\rm Tr}_{\bar{\rm S}}\{\hat{U}(t,t_0)\hat{\rho}^{t_0}_{\rm tot}(t_0)\hat{U}^\dagger(t,t_0)\}.
\end{equation}
Note that since Eq.~\eqref{eq:initst2} provides the initial condition, we only consider times $t\geq t_0$.

If $\Sigma$ is independent of our choice of $t_0$, then it can be shown that $\Sigma\geq 0$ at all times since
\begin{equation}
\label{eq:secondlaw2}
\begin{aligned}
\Sigma(t) &\simeq \frac{S\left[{\textstyle \hat{\rho}^{t_0}_{\rm tot}(t+dt)||\hat{\rho}_{\rm S}^{t_0}(t+dt)\bigotimes_\alpha\hat{\tau}_{\beta_\alpha,\mu_\alpha}}\right]-S\left[{\textstyle \hat{\rho}^{t_0}_{\rm tot}(t)||\hat{\rho}_{\rm S}^{t_0}(t)\bigotimes_\alpha\hat{\tau}_{\beta_\alpha,\mu_\alpha}}\right]}{dt}\\&= \frac{S\left[{\textstyle \hat{\rho}^{t}_{\rm tot}(t+dt)||\hat{\rho}_{\rm S}^{t}(t+dt)\bigotimes_\alpha\hat{\tau}_{\beta_\alpha,\mu_\alpha}}\right]}{dt}\geq 0,
\end{aligned}
\end{equation}
where we have chosen $t=t_0$ in the second line.

As we will see in the next section, in the limit of infinitely large and memory-less reservoirs which couple weakly to the system, no observables depend on the time $t_0$ at which we assume system and reservoirs to be uncorrelated. This ensures that for \textit{classical} reservoirs, the entropy production rate is positive at all times and suggests the following picture: In reality, the state never truly factorizes as in Eq.~\eqref{eq:initst2}. However, at all times it is very close to it. At all times, we can thus assume it to be of the form given in Eq.~\eqref{eq:initst2}. However, once we fix the state to be of this form for a given time, we also need to allow for (small) deviations of this form. These deviations encode the energy flows and determine the entropy production according to Eq.~\eqref{eq:entprod}.

{\small
	\paragraph{Exercises}
	
	\paragraph{2.1 Entropy production in a thermal bath (2pt)}
	Consider a system which is in a thermal state at all times, but has time-dependent temperature and chemical potential
	\begin{equation}
	\label{eq:thermtex}
	\hat{\tau}_{\beta_t,\mu_t}=\frac{e^{-\beta_t(\hat{H}-\mu_t\hat{N})}}{Z_t}.
	\end{equation}
	Physically, this is obtained if the system has internal interactions which result in a very fast thermalization. We consider the scenario where the temperature and the chemical potential variations are small and write
	\begin{equation}
	\label{eq:betamutex}
	\beta_t=\beta\left[1+\delta\cdot f_\beta (t)\right],\hspace{2cm}\mu_t=\mu\left[1+\delta\cdot f_\mu(t)\right].
	\end{equation}
	Show that for $\delta\ll1$
	\begin{equation}
	\label{eq:entthermex}
	\partial_tS_{\rm vN}[\hat{\tau}_{\beta_t,\mu_t}]=\beta\partial_t\langle\hat{H}-\mu\hat{N}\rangle+\mathcal{O}(\delta^2).
	\end{equation}
	\paragraph{Hint:} You can make use of the relation
	\begin{equation}
	\label{eq:hint21}
	{\rm Tr}\left\{\partial_te^{\hat{A}(t)}\right\}={\rm Tr}\left\{e^{\hat{A}(t)}\partial_t\hat{A}(t)\right\}.
	\end{equation}

	\paragraph{2.2 Dissipated work (3pt)}
	Consider the scenario where a system coupled to a single bath is driven by a time-dependent Hamiltonian. We consider a system which can only exchange energy with the reservoir (i.e., we set $\mu=0$). Further consider a process which starts and ends with the system being in thermal equilibrium, i.e.
	\begin{equation}
	\label{eq:initfinstate}
	\hat{\rho}_{\rm S}(0)=\frac{e^{-\beta\hat{H}_{\rm S}(0)}}{Z(0)}=e^{-\beta[\hat{H}_{\rm S}(0)-F_{\rm S}(0)]},\hspace{1.5cm}\hat{\rho}_{\rm S}(\tau)=\frac{e^{-\beta\hat{H}_{\rm S}(\tau)}}{Z(\tau)}=e^{-\beta[\hat{H}_{\rm S}(\tau)-F_{\rm S}(\tau)]},
	\end{equation}
	where we introduced the free energy of the system 
	\begin{equation}
	\label{eq:freeenergy}
	F_{\rm S}={\rm Tr}\{\hat{H}_{\rm S}\hat{\rho}_{\rm S}\}-k_{\rm B} TS_{\rm vN}[\hat{\rho}_{\rm S}].
	\end{equation}
	Additionally, we assume the initial total state to be a product state between system and bath
	\begin{equation}
	\label{eq:initstate}
	\hat{\rho}_{\rm tot}(0)=\hat{\rho}_{\rm S}(0)\otimes\hat{\tau}_\beta.
	\end{equation}
	Note that the final state is in general \textit{not} a product state. Equation \eqref{eq:initfinstate} only restricts the reduced state of the system. First, show that for a thermal state
	\begin{equation}
	F_{\rm S}=-k_{\rm B} T\ln Z.
	\end{equation}
	
	Using the first and second laws of thermodynamics, show that the amount of work spent for this process is at least as big as the difference of free energy between final and initial state of the system, i.e.,
	\begin{equation}
	\label{eq:secondlawwork}
	W_{\rm S} = \int_0^\tau dtP_{\rm S}(t)\geq \Delta F_{\rm S}=F_{\rm S}(\tau)-F_{\rm S}(0).
	\end{equation}
	
	Further show that the equality is obtained for a quasi-static process, where the reduced state of the system is an equilibrium state at all times
	\begin{equation}
	\label{eq:statworkpr}
	\hat{\rho}_{\rm S}(t)={\rm Tr}_{\bar{\rm S}}\{\hat{\rho}_{\rm tot}\}=e^{-\beta[\hat{H}_{\rm S}(t)-F_{\rm S}(t)]}.
	\end{equation}
	Summarize the assumptions which went into the obtained results.
	\paragraph{Hint:} For the second part, you can make use of the relation
	\begin{equation}
	\label{eq:derexphint}
	{\rm Tr}\left\{\partial_te^{\hat{A}(t)}\right\}={\rm Tr}\left\{e^{\hat{A}(t)}\partial_t\hat{A}(t)\right\}.
	\end{equation}
	
	\paragraph{2.3 Invariance of the von Neumann entropy (1pt)}
	Show that the von Neumann entropy does not change under unitary dynamics. I.e., show that
	\begin{equation}
	\partial_tS_{\rm vN}[\hat{\rho}(t)]=0,
	\end{equation}
	for
	\begin{equation}
	\hat{U}(t)\hat{\rho}(0)\hat{U}^\dagger(t),\hspace{2cm}\hat{U}(t)\hat{U}^\dagger(t)=\mathbb{I}.
	\end{equation}}

\section{Markovian master equations}
In this section, we consider Markovian master equations as a description for the reduced state of the system. Markovianity implies that no memory effects of the bath are taken into account (i.e., a particle that is emitted from the system will not be reabsorbed by the system before losing all memory of the emission event). In principle, memory effects are always present but if the coupling between system and bath is weak, these effects can often safely be ignored. There are numerous references that discuss Markovian master equations going substantially beyond these notes, see for instance Refs.~\cite{breuer:book,rivas:book,schaller:book,carmichael:book}

\begin{tcolorbox}
	\paragraph{Operators and superoperators} In quantum mechanics, an operator refers to a linear operator that transforms a vector in a Hilbert space of dimension $d$, $\mathscr{H}_d$, into another vector in the same Hilbert space, i.e.,
	\[\hat{O}|\psi\rangle\in \mathscr{H}_d \hspace{.5cm}\forall\hspace{.5cm}|\psi\rangle\in \mathscr{H}_d. \]
	An operator can itself be interpreted as an element of a vector space with dimension $d^2$, the space of linear operators sometimes called Liouville space $\mathscr{L}_{d^2}$. Introducing the scalar product $\langle \hat{A},\hat{B}\rangle={\rm Tr}\{\hat{A}^\dagger\hat{B}\}$, this vector space becomes a Hilbert space (a Hilbert space is a complex vector space equipped with a scalar product). A superoperator is a linear operator that acts on the elements of the Liouville space $\mathscr{L}_{d^2}$. 
	superoperators thus act on operators in the same way that operators act on quantum states and we have
	\[\mathcal{S}\hat{A}\in\mathscr{L}_{d^2}\hspace{.5cm}\forall\hspace{.5cm}\hat{A}\in\mathscr{L}_{d^2}.\]
	For finite $d$, one can explicitly write $\hat{A}$ as a column vector in $\mathbb{C}^{d^2}$ such that each entry of this column vector equals a matrix element of $\hat{A}$. Any superoperator can then be written as a $d^2\times d^2$ dimensional matrix \cite{puri:book}.
	
\end{tcolorbox}

\begin{tcolorbox}
	\paragraph{UDMs, Markovianity, and the GKLS form} A universal dynamical map (UDM) $\mathcal{E}$ is a linear map which transforms a density matrix into another density matrix. Furthermore, it is independent of the density matrix it acts upon. In its most general form, a UDM can be written as
	\[\mathcal{E}\hat{\rho}=\sum_j\hat{K}_j\hat{\rho}\hat{K}_j^\dagger,\hspace{1.5cm}\sum_j\hat{K}_j^\dagger\hat{K}_j=\mathbb{I}.\]
	
	We say that a system obeys Markovian time-evolution if it is described by a \textit{divisible} UDM, i.e.,
	\[\hat{\rho}(t)=\mathcal{E}_{t,t_0}\hat{\rho}=\mathcal{E}_{t,t_1}\mathcal{E}_{t_1,t_0}\hat{\rho},\]
	For any intermediate time $t_1$. Finally, we note that a differential equation is a Markovian master equation (i.e., results in Markovian time-evolution) if and only if it can be written in the form
	\begin{equation}
	\label{eq:gkls}
	\partial_t\hat{\rho}(t)=-i[\hat{H}(t),\hat{\rho}(t)]+\sum_k\gamma_k(t)\left[\hat{S}_k(t)\hat{\rho}(t)\hat{S}_k^\dagger(t)-\frac{1}{2}\{\hat{S}^\dagger_k(t)\hat{S}_k(t),\hat{\rho}(t)\}\right],
	\end{equation}
	where $\hat{H}(t)$ is Hermitian and $\gamma_k(t)\geq 0$. For a proof, see Ref.~\cite{rivas:book}. This form of the master equation is also called GKLS form, after Gorini, Kosakowski, Sudarshan \cite{gorini:1976}, and Linblad \cite{lindblad:1976} who considered the time-independent case. 
\end{tcolorbox}

Here we provide a general derivation of a Markovian master equation. We use the Nakajima-Zwanzig projection operator approach \cite{nakajima:1958,zwanzig:1960} following Ref.~\cite{rivas:2010}. To this end, we introduce the superoperators (see box)
\begin{equation}
\label{eq:nzso}
\mathcal{P}\hat{\rho}_{\rm tot}={\rm Tr}_{\bar{S}}\left\{\hat{\rho}_{\rm tot}\right\}\bigotimes_\alpha\hat{\tau}_{\beta_\alpha,\mu_\alpha}=\hat{\rho}_S\bigotimes_\alpha\hat{\tau}_{\beta_\alpha,\mu_\alpha},\hspace{1.5cm}\mathcal{Q}=\mathcal{I}-\mathcal{P},
\end{equation} 
where $\mathcal{I}$ denotes the identity. Note that these are projectors as $\mathcal{P}^2=\mathcal{P}$. Further, note that we are interested in the time-evolution of $\mathcal{P}\hat{\rho}_{\rm tot}(t)$, which provides us with an effective description of the form discussed in Sec.~\ref{sec:entropy}. We consider the general scenario discussed in Sec.~\ref{sec:general}, i.e., the Hamiltonian reads
\begin{equation}
\label{eq:htot2}
\hat{H}_{\rm tot}(t)=\hat{H}_{\rm S}(t)+\sum_{\alpha}\left(\hat{H}_\alpha+\hat{H}_{\alpha \rm S}\right).
\end{equation}
We now go to an interaction picture
\begin{equation}
\label{eq:interact}
\tilde{\rho}_{\rm tot}(t)=\hat{V}^\dagger(t)\hat{\rho}_{\rm tot}(t)\hat{V}(t),
\end{equation}
where
\begin{equation}
\label{eq:interact2}
\hat{V}(t)=\mathcal{T}e^{-i\int_{0}^{t}dt'[\hat{H}_{\rm S}(t')+\sum_{\alpha}\hat{H}_\alpha]},
\end{equation}
with $\mathcal{T}$ denoting time-ordering. In the interaction picture, the time-evolution of the total density matrix is determined by
\begin{equation}
\label{eq:vneutotint}
\partial_t\tilde{\rho}_{\rm tot}(t)=-i\sum_{\alpha}\left[\tilde{H}_{\alpha\rm S}(t),\tilde{\rho}_{\rm tot}(t)\right]=\mathcal{V}(t)\tilde{\rho}_{\rm tot}(t),
\end{equation}
where we used Eq.~\eqref{eq:rhototu} as well as
\begin{equation}
\partial_t\hat{V}(t)=-i\left[\hat{H}_{\rm S}(t)+\sum_{\alpha}\hat{H}_\alpha\right]\hat{V}(t),\hspace{1.5cm}\partial_t\hat{V}^\dagger(t)=i\hat{V}^\dagger(t)\left[\hat{H}_{\rm S}(t)+\sum_{\alpha}\hat{H}_\alpha\right],
\end{equation}
and the coupling Hamiltonian in the interaction picture is obtained analogously to Eq.~\eqref{eq:interact}. Finally, we have expressed the commutator in Eq.~\eqref{eq:vneutotint} with the help of the superoperator $\mathcal{V}$. In the following, we will assume $\mathcal{P}\mathcal{V}(t)\mathcal{P}=0$. This is not a restriction as it can always be obtained by adding some terms to $\hat{H}_{\rm S}$ and subtracting them from $\hat{H}_{\alpha\rm S}$ \cite{schaller:book}. We can then write
\begin{equation}
\label{eq:dtpchi}
\partial_t\mathcal{P}\tilde{\rho}_{\rm tot}(t)=\mathcal{P}\mathcal{V}(t)\tilde{\rho}_{\rm tot}(t)=\mathcal{P}\mathcal{V}(t)\mathcal{Q}\tilde{\rho}_{\rm tot}(t),
\end{equation}
\begin{equation}
\label{eq:dtqchi}
\partial_t\mathcal{Q}\tilde{\rho}_{\rm tot}(t)=\mathcal{Q}\mathcal{V}(t)\tilde{\rho}_{\rm tot}(t)=\mathcal{Q}\mathcal{V}(t)\mathcal{P}\tilde{\rho}_{\rm tot}(t)+\mathcal{Q}\mathcal{V}(t)\mathcal{Q}\tilde{\rho}_{\rm tot}(t),
\end{equation}
where we used $\mathcal{P}+\mathcal{Q}=\mathcal{I}$. The formal solution to the second equation is given by
\begin{equation}
\label{eq:qchifsol}
\mathcal{Q}\tilde{\rho}_{\rm tot}(t) = \mathcal{G}(t,0)\mathcal{Q}\tilde{\rho}_{\rm tot}(0)+\int_{0}^{t}ds\mathcal{G}(t,s)\mathcal{Q}\mathcal{V}(s)\mathcal{P}\tilde{\rho}_{\rm tot}(s),
\end{equation}
where we introduced the propagator
\begin{equation}
\label{eq:propg}
\mathcal{G}(t,s) =\mathcal{T}e^{\int_{s}^{t}\mathcal{Q}\mathcal{V}(\tau)d\tau}.
\end{equation}

We now assume factorizing initial conditions
\begin{equation}
\tilde{\rho}_{\rm tot}(0)=\hat{\rho}_S(0)\bigotimes_\alpha\hat{\tau}_{\beta_\alpha,\mu_\alpha},
\end{equation}
such that $\mathcal{P}\tilde{\rho}_{\rm tot}(0)=\tilde{\rho}_{\rm tot}(0)$ and $\mathcal{Q}\tilde{\rho}_{\rm tot}(0)=0$. Inserting Eq.~\eqref{eq:qchifsol} into Eq.~\eqref{eq:dtpchi}, we find
\begin{equation}
\label{eq:dtpchi2}
\partial_t\mathcal{P}\tilde{\rho}_{\rm tot}(t)=\int_{0}^{t}ds\mathcal{P}\mathcal{V}(t)\mathcal{G}(t,s)\mathcal{Q}\mathcal{V}(s)\mathcal{P}\tilde{\rho}_{\rm tot}(s).
\end{equation}
This expression is still exact (for the given initial conditions).

\subsection{Born-Markov approximations}
We now make a weak coupling approximation. If the coupling between system and baths are proportional to $r$, we find
\begin{equation}
\label{eq:dtpchi3}
\partial_t\mathcal{P}\tilde{\rho}_{\rm tot}(t)=\mathcal{P}\int_{0}^{t}ds\mathcal{V}(t)\mathcal{V}(s)\mathcal{P}\tilde{\rho}_{\rm tot}(s)+\mathcal{O}(r^3),
\end{equation}
where we again used $\mathcal{P}\mathcal{V}(t)\mathcal{P}=0$. The last equation implies
\begin{equation}
\label{eq:bornapp}
\partial_t\tilde{\rho}_{\rm S}(t)=-\int_{0}^{t}ds\sum_{\alpha}{\rm Tr}_{\rm \bar{S}}\left\{\left[\tilde{H}_{\alpha\rm S}(t),\left[\tilde{H}_{\alpha\rm S}(t-s),\tilde{\rho}_{\rm S}(t-s)\otimes_\alpha\hat{\tau}_{\beta_\alpha,\mu_\alpha}\right]\right]\right\},
\end{equation}
where we substituted $s\rightarrow t-s$ and we made use of ${\rm Tr}_{\rm \bar{S}}\{\tilde{H}_{\alpha\rm S}(t)\tilde{H}_{\beta\rm S}(s)\tilde{\rho}_{\rm S}(s)\otimes_\alpha\hat{\tau}_{\beta_\alpha,\mu_\alpha}\}=0$ for $\alpha\neq\beta$. This is similar to the assumption $\mathcal{P}\mathcal{V}(t)\mathcal{P}=0$ and can always be ensured by an appropriate redefinition of the terms in the Hamiltonian. We note that Eq.~\eqref{eq:bornapp} is often obtained by assuming that $\tilde{\rho}_{\rm tot}(t)=\tilde{\rho}_{\rm S}(t)\otimes_\alpha\hat{\tau}_{\beta_\alpha,\mu_\alpha}$ at all times. Here we do not make such an assumption. In agreement with the discussion in the previous section, we consider $\tilde{\rho}_{\rm S}(t)\otimes_\alpha\hat{\tau}_{\beta_\alpha,\mu_\alpha}$ to be an effective description, which only keeps track of the degrees of the system and neglects changes in the bath state as well as correlations between system and bath.

In addition to the weak coupling approximation, we now make a Markov approximation. To this end, we assume that the integrand in Eq.~\eqref{eq:bornapp} decreases on a time-scale $\tau_{\rm B}$ (the bath-correlation time, more on this below). If this time-scale is short enough, which is the case for large, memory-less baths, we can assume $\tilde{\rho}_{\rm S}$ to approximately remain constant and replace its time-argument in Eq.~\eqref{eq:bornapp} by $t$. Furthermore, using the same argumentation, we can extend the integral to infinity obtaining
\begin{equation}
\label{eq:bornmarkov}
\partial_t\tilde{\rho}_{\rm S}(t)=-\int_{0}^{\infty}ds\sum_{\alpha}{\rm Tr}_{\rm \bar{S}}\left\{\left[\tilde{H}_{\alpha\rm S}(t),\left[\tilde{H}_{\alpha\rm S}(t-s),\tilde{\rho}_{\rm S}(t)\otimes_\alpha\hat{\tau}_{\beta_\alpha,\mu_\alpha}\right]\right]\right\}.
\end{equation}
This equation is Markovian, i.e., it is local in time and does not depend explicitly on the initial conditions. However, it is not in GKLS form (see box) and does not in general preserve the positivity of the density matrix. The approximations that result in Eq.~\eqref{eq:bornmarkov} are usually called the \textit{Born-Markov} approximations. For a more formal application of these approximations, see Refs.~\cite{davies:1974,davies:1976}. Note that under the Born-Markov approximations, the effect induced by different baths is additive.

To make progress, we write the coupling Hamiltonian in the general form
\begin{equation}
\label{eq:couptens}
\hat{H}_{\alpha\rm S} = \sum_k \hat{S}_{\alpha,k}\otimes\hat{B}_{\alpha,k}=\sum_k\hat{S}^\dagger_{\alpha,k}\otimes\hat{B}^\dagger_{\alpha,k},
\end{equation}
where we used the Hermiticity of $\hat{H}_{\alpha\rm S}$ in the second equality.
We note that the operators $\hat{S}_{\alpha,k}$ and $\hat{B}_{\alpha,k}$ are not necessarily Hermitian. Furthermore, we note that the tensor product structure is not provided for fermions, where operators on the bath and on the system may \textit{anti}-commute. However, using a Jordan-Wigner transform, one can map the fermionic system onto a spin system where such a tensor-product structure is provided \cite{schaller:book}. After tracing out the bath, the spin operators can then be replaced by fermionic operators again. Inserting Eq.~\eqref{eq:couptens} into Eq.~\eqref{eq:bornmarkov}, we find after some algebra
\begin{equation}
\label{eq:bornmarkov2}
\begin{aligned}
\partial_t\tilde{\rho}_{\rm S}(t)=\sum_{\alpha}\sum_{k,k'}\int_{0}^{\infty}&ds\left\{C^\alpha_{k,k'}(s)\left[\tilde{S}_{\alpha,k'}(t-s)\tilde{\rho}_{\rm S}(t)\tilde{S}^\dagger_{\alpha,k}(t)-\tilde{S}^\dagger_{\alpha,k}(t)\tilde{S}_{\alpha,k'}(t-s)\tilde{\rho}_{\rm S}(t)\right]\right.\\&\left.+C^\alpha_{k,k'}(-s)\left[\tilde{S}_{\alpha,k'}(t)\tilde{\rho}_{\rm S}(t)\tilde{S}^\dagger_{\alpha,k}(t-s)-\tilde{\rho}_{\rm S}(t)\tilde{S}^\dagger_{\alpha,k}(t-s)\tilde{S}_{\alpha,k'}(t)\right]\right\},
\end{aligned}
\end{equation}
where we introduced the bath correlation functions
\begin{equation}
\label{eq:bathcorr}
C^\alpha_{k,k'}(s)={\rm Tr}\left\{\tilde{B}^\dagger_{\alpha,k}(s)\hat{B}_{\alpha,k'}\hat{\tau}_{\beta_\alpha,\mu_\alpha}\right\},
\end{equation}
and we used $[C^\alpha_{k,k'}(s)]^*=C^\alpha_{k',k}(-s)$. These bath correlation functions are usually peaked around $s=0$ and decay over the time-scale $\tau_{\rm B}$ (indeed, this is how $\tau_{\rm B}$ is defined). If this time-scale is short, the integrand in Eq.~\eqref{eq:bornmarkov2} decays quickly and the Markov assumption performed above is justified. Note that it is important that this approximation is made in the interaction picture, where $\tilde{\rho}_{\rm S}$ varies slowly (in the Schr\"odinger picture, $\hat{\rho}_{\rm S}$ tends to oscillate with frequencies given by the differences of the eigenvalues of $\hat{H}_{\rm S}$).

\subsection{Single-component systems}
As mentioned above, Eq.~\eqref{eq:bornmarkov2} does not guarantee preservation of positivity (i.e., for certain initial conditions and structures of the Hamiltonian, the time evolution may result in density matrices which have negative eigenvalues). There is however an important class of Hamiltonians for which the Born-Markov approximations are sufficient to obtain a master equation in the GKLS form (see box) which preserves positivity. Such systems are here called \textit{single-component systems}. Consider a coupling Hamiltonian of the form of Eq.~\eqref{eq:couptens} which fulfills the condition (for a time-independent system Hamiltonian)
\begin{equation}
\label{eq:sjump}
[\hat{S}_{\alpha,k},\hat{H}_{\rm S}]=\omega_{\alpha,k}\hat{S}_{\alpha,k}.
\end{equation}
In this case, $\hat{S}_{\alpha,k}$ is a ladder operator that, when acting on an eigenstate of the Hamiltonian, returns another eigenstate of the Hamiltonian with an energy that is lower by $\omega_{\alpha,k}$. In this case, one can show using the Baker-Campbell-Hausdorff formula
\begin{equation}
\label{eq:sint}
\tilde{S}_{\alpha,k}(t)=e^{i\hat{H}_{\rm S}t}\hat{S}_{\alpha,k}e^{-i\hat{H}_{\rm S}t}=e^{-i\omega_{\alpha,k}t}\hat{S}_{\alpha,k}.
\end{equation}
We will furthermore consider the restriction
\begin{equation}
\label{eq:restb}
C^\alpha_{k,k'}(s)\propto \delta_{k,k'}.
\end{equation}
In this case, the master equation reduces to
\begin{equation}
\label{eq:mastersingcomp}
\partial_t\hat{\rho}_{\rm S}=-i[\hat{H}_{\rm S},\hat{\rho}_{\rm S}]+\sum_{\alpha}\sum_{k}\left\{\gamma_{k}^\alpha\left[\hat{S}_{\alpha,k}\hat{\rho}_{\rm S}\hat{S}^\dagger_{\alpha,k}-\frac{1}{2}\{\hat{S}^\dagger_{\alpha,k}\hat{S}_{\alpha,k},\hat{\rho}_{\rm S}\}\right]-i\Delta_{k}^\alpha[\hat{S}^\dagger_{\alpha,k}\hat{S}_{\alpha,k},\hat{\rho}_{\rm S}]\right\},
\end{equation}
where we introduced
\begin{equation}
\label{eq:gammas}
\gamma_{k}^\alpha=2{\rm Re}\left\{\int_0^\infty dse^{i\omega_{\alpha,k} s} C^\alpha_{k,k}(s)\right\},
\end{equation}
\begin{equation}
\label{eq:deltas}
\Delta_{k}^\alpha={\rm Im}\left\{\int_0^\infty dse^{i\omega_{\alpha,k} s} C^\alpha_{k,k}(s)\right\}.
\end{equation}
Note that the last term in Eq.~\eqref{eq:mastersingcomp} modifies the Hamiltonian (i.e., the unitary part of the time evolution). This term is called the Lamb shift and is often neglected. Since $\gamma_{k}^\alpha\geq0$ (this follows from Bochners theorem and the fact that $\gamma$ can be written as the Fourier transform of a positive definite function), the master equation in Eq.~\eqref{eq:mastersingcomp} is in GKLS form and thus preserves positivity. Note that in order to go back to the Schr\"odinger picture, we used
\begin{equation}
\label{eq:backschro}
\partial_t\hat{\rho}_{\rm S}(t)=-i[\hat{H}_{\rm S},\hat{\rho}_{\rm S}(t)]+e^{-i\hat{H}_{\rm S}t}\partial_t\tilde{\rho}_{\rm S}(t)e^{i\hat{H}_{\rm S}t}.
\end{equation}

\subsection{Additional approximations}
\label{sec:addapp}
For many interesting scenarios, the Born-Markov approximations are not sufficient to guarantee positivity of the density matrix. Here we consider three different approaches to obtain a master equation in GKLS form from Eq.~\eqref{eq:bornmarkov2}.

\subsubsection{Secular approximation} 
The secular approximation is the most well known approach for obtaining a master equation in GKLS form and can be found in many text-books (see for instance Ref.~\cite{breuer:book}). To perform the secular approximation, we write the system operators in the interaction picture as
\begin{equation}
\label{eq:sfour}
\tilde{S}_{\alpha,k}(t) = \sum_{a,b}e^{i (E_a-E_b)t}|E_a\rangle\langle E_a|\hat{S}_{\alpha,k}|E_b\rangle\langle E_b|=\sum_j e^{-i\omega_j t}\hat{S}^j_{\alpha,k},
\end{equation}
where we introduced the eigenstates of the system Hamiltonian $\hat{H}_{\rm S}|E_a\rangle = E_a|E_a\rangle$ and we grouped all gaps of equal sizes into $\omega_j$, ensuring $\omega_i\neq\omega_j$ for $i\neq j$. The secular approximation then consists of inserting Eq.~\eqref{eq:sfour} into Eq.~\eqref{eq:bornmarkov2} and dropping all terms which oscillate with frequency $\omega_i-\omega_j$ for $i\neq j$. The resulting master equation is in GKLS form and reads
\begin{equation}
\label{eq:secular}
\begin{aligned}
\partial_t\hat{\rho}_{\rm S}=-i[\hat{H}_{\rm S},\hat{\rho}_{\rm S}]+\sum_{\alpha}\sum_{k,k'}\sum_j&\left\{\Gamma_{k,k'}^\alpha(\omega_j)\left[\hat{S}^j_{\alpha,k'}\hat{\rho}_{\rm S}\left(\hat{S}^j_{\alpha,k}\right)^\dagger-\frac{1}{2}\left\{\left(\hat{S}^j_{\alpha,k}\right)^\dagger\hat{S}^j_{\alpha,k'},\hat{\rho}_{\rm S}\right\}\right]\right.\\&\left.
-i\Delta^\alpha_{k,k'}(\omega_j)\left[\left(\hat{S}^j_{\alpha,k}\right)^\dagger\hat{S}^j_{\alpha,k'},\hat{\rho}_{\rm S}\right]\right\},
\end{aligned}
\end{equation}
where we introduced
\begin{equation}
\label{eq:biggam}
\Gamma_{k,k'}^\alpha(\omega_j) = \int_{-\infty}^{\infty}dse^{i\omega_js}C_{k,k'}^\alpha(s),\hspace{1cm}\Delta^\alpha_{k,k'}(\omega_j)=-\frac{i}{2}\int_{-\infty}^{\infty}dse^{i\omega_j s}{\rm sign}(s)C_{k,k'}^\alpha(s).
\end{equation}
Equation \eqref{eq:secular} is in GKLS form because $\Delta^\alpha_{k,k'}(\omega_j)=[\Delta^\alpha_{k',k}(\omega_j)]^*$ and it can be shown that $\Gamma_{k,k'}^\alpha(\omega_j)$ is a positive semi-definite matrix \cite{schaller:book}.

The secular approximation is valid as long as the terms that oscillate with frequency $\omega_j-\omega_{j'}$ for $j\neq j'$ can be neglected. This is the case if $\tilde{\rho}_{\rm S}$ is approximately constant during many such oscillations. Let us call the time-scale over which $\tilde{\rho}_{\rm S}$ varies $\tau_{\rm S}$. The secular approximation is then justified when $|(\omega_j-\omega_{j'})|\tau_{\rm S}\gg1$ for all $j\neq j'$ and is thus expected to work well for systems which have no small gaps in $\hat{H}_{\rm S}$. As we will see, many thermal machines consist of multiple parts that are weakly coupled and have small energy gaps induced by the weak coupling. In such systems, the secular approximation is expected to break down. Note that in order to obtain the jump operators $\hat{S}^j_{\alpha,k'}$, the Hamiltonian $\hat{H}_{\rm S}$ needs to be diagonalized first. This implies that even obtaining the master equation may be a formidable task.

A particularly appealing feature of the secular approximation is the fact that the second law of thermodynamics (i.e., $\Sigma\geq0$) is ensured by construction \cite{spohn:1978,alicki:1979}.
We note that in the secular approximation, for a non-degenerate system Hamiltonian, the populations decouple from the coherences. Concretely, this means that in the energy-eigenbasis, the time-evolution of the off-diagonal terms of the density matrix are independent of the diagonal elements and vice versa. Usually, the off-diagonal elements tend to zero, i.e., the bath suppresses any coherences between the energy eigenstates. While this may be the case, there are many situations of interest where coherences between energy eigenstates are important in the presence of thermal baths and the secular approximation is no longer justified \cite{hofer:2017njp,mitchison:2018,gonzalez:2017,seah:2018,kirsanskas:2018}.

\subsubsection{Second Markov approximation}
The second Markov approximation is another popular approach (often called the singular coupling limit \cite{breuer:book,schaller:book}). In this approach, we assume that $\tilde{S}_{\alpha,k}(t)$ in Eq.~\eqref{eq:bornmarkov2} varies on a time-scale that is much smaller than $\tau_{\rm B}$. In complete analogy to the first Markov approximation, we can then replace $\tilde{S}_{\alpha,k}(t-s)\rightarrow\tilde{S}_{\alpha,k}(t)$ in the integral of Eq.~\eqref{eq:bornmarkov2}. This results in
\begin{equation}
\label{eq:secondmarkov}
\begin{aligned}
\partial_t\hat{\rho}_{\rm S}=-i[\hat{H}_{\rm S},\hat{\rho}_{\rm S}]+\sum_{\alpha}\sum_{k,k'}&\left\{\Gamma_{k,k'}^\alpha\left[\hat{S}_{\alpha,k'}\hat{\rho}_{\rm S}\hat{S}_{\alpha,k}^\dagger-\frac{1}{2}\left\{\hat{S}_{\alpha,k}^\dagger\hat{S}_{\alpha,k'},\hat{\rho}_{\rm S}\right\}\right]\right.\\&\left.
-i\Delta^\alpha_{k,k'}\left[\hat{S}_{\alpha,k}^\dagger\hat{S}_{\alpha,k'},\hat{\rho}_{\rm S}\right]\right\},
\end{aligned}
\end{equation}
which is in GKLS form. From Eq.~\eqref{eq:sint}, we see that $\tilde{S}_{\alpha,k}(t)$ may oscillate quickly. However, these fast oscillations can often be removed from $\tilde{S}_{\alpha,k}(t)$ (and transferred to $\tilde{B}_{\alpha,k}(t)$) by going to a rotating frame. This substantially increases the reliability of the second Markov approximation. An important advantage of the second Markov approximation over the secular approximation is the fact that it can be performed without explicitly diagonalizing the system Hamiltonian.

The second Markov approximation is particularly useful for multiple single-component systems that are weakly coupled. In this case, the approximation results in the same master equation one obtains by deriving the dissipative terms for each single-component system in isolation and adding them up to describe the coupled system. Due to this local structure, the second Markov approximation provides a \textit{local} approach, in contrast to the secular approximation which provides a \textit{global} approach.

We note that in contrast to the secular approximation, the second law is not guaranteed to hold under the second Markov approximation. However, as discussed in Sec.~\ref{sec:secondlaw}, the second law should hold as long as the master equation produces a faithful description of the system density matrix. Any violation of the second law is therefore due to terms that are neglected in the derivation of the Markovian master equation. These are small as long as the Markovian master equation provides an appropriate description of the system. Note however that a violation of the second law can still arise when $\Sigma$ itself is of the order of the neglected terms.

\subsubsection{Position and Energy Resolved Lindblad (PERLind) approach} 
Here we discuss the approach that was introduced by Kir{\v s}anskas, Francki\'e and Wacker in Ref.~\cite{kirsanskas:2018}. We follow the more formal motivation of this approach given in the supplemental material of Ref.~\cite{ptaszynski:2019}. First, we use the expansion of the jump operators in Eq.~\eqref{eq:sfour} to rewrite Eq.~\eqref{eq:bornmarkov2} as
\begin{equation}
\label{eq:bornmarkov3}
\begin{aligned}
\partial_t\hat{\rho}_{\rm S}=-i[\hat{H}_{\rm S},\hat{\rho}_{\rm S}]+\sum_{\alpha}\sum_{k,k'}\sum_{j,j'}&\left\{\mathcal{G}_{k,k'}^\alpha(\omega_j,\omega_{j'})\left[\hat{S}^{j}_{\alpha,k'}\hat{\rho}_{\rm S}\left(\hat{S}^{j'}_{\alpha,k}\right)^\dagger-\frac{1}{2}\left\{\left(\hat{S}^{j'}_{\alpha,k}\right)^\dagger\hat{S}^j_{\alpha,k'},\hat{\rho}_{\rm S}\right\}\right]\right.\\&\left.
-i\mathcal{D}^\alpha_{k,k'}(\omega_j,\omega_{j'})\left[\left(\hat{S}^{j'}_{\alpha,k}\right)^\dagger\hat{S}^j_{\alpha,k'},\hat{\rho}_{\rm S}\right]\right\},
\end{aligned}
\end{equation}
where we introduced
\begin{equation}
\label{eq:paragen}
\begin{aligned}
&\mathcal{G}_{k,k'}^\alpha(\omega_j,\omega_{j'})=\frac{1}{2}\left[\Gamma_{k,k'}^\alpha(\omega_{j'})+\Gamma_{k,k'}^\alpha(\omega_{j})\right]+i\left[\Delta_{k,k'}^\alpha(\omega_{j'})-\Delta_{k,k'}^\alpha(\omega_{j})\right],\\
&\mathcal{D}_{k,k'}^\alpha(\omega_j,\omega_{j'})=\frac{1}{2}\left[\Delta_{k,k'}^\alpha(\omega_{j'})-\Delta_{k,k'}^\alpha(\omega_{j})\right]-\frac{i}{4}\left[\Gamma_{k,k'}^\alpha(\omega_{j'})-\Gamma_{k,k'}^\alpha(\omega_{j})\right],
\end{aligned}
\end{equation}
with coefficients given in Eq.~\eqref{eq:biggam}. This equation is not in GKLS form. The secular approximation is obtained from this equation by dropping all terms where $j\neq j'$. As mentioned above, this is justified as long as $|(\omega_j-\omega_{j'})|\tau_{\rm S}\gg1$. The second Markov approximation is obtained by assuming that $\Gamma_{k,k'}^\alpha$ and $\Delta_{k,k'}^\alpha$ can be taken to be independent of $\omega$. This is justified as long as $|(\omega_j-\omega_{j'})|\tau_{\rm B}\ll1$. This can be seen by noting that the coefficients in Eq.~\eqref{eq:bornmarkov3} are determined by the Fourier transform of the bath correlation functions. Since the bath correlation functions decay on a time-scale $\tau_{\rm B}$, their Fourier transforms are constant over the energy scale $1/\tau_{B}$.

To obtain the PERLind master equation, we group the energy gaps into two cases:
\begin{equation}
|(\omega_j-\omega_{j'})|\tau_{\rm S}\gg1,
\end{equation} 
and
\begin{equation}
|(\omega_j-\omega_{j'})|\tau_{\rm S}\lesssim1 \hspace{.5cm}\Rightarrow\hspace{.5cm}|(\omega_j-\omega_{j'})|\tau_{\rm B}\ll1,
\end{equation}
where we used $\tau_{\rm B}\ll\tau_{\rm S}$ which is required in order for the Born-Markov approximations to be justified. If all energy gaps fulfill the first condition, the secular approximation is justified. If all fulfill the second condition, the second Markov approximation is justified. In the spirit of the secular approximation, we can modify the terms fulfilling the first condition since they do not contribute to the dynamics anyway. In the spirit of the second Markov approximation, we can assume $\Gamma_{k,k'}^\alpha(\omega_j)\simeq \Gamma_{k,k'}^\alpha(\omega_{j'})$ and similarly for $\Delta_{k,k'}^\alpha$ for all energy gaps which fulfill the second condition. It is therefore justified to perform the following substitutions in \textit{all} terms of Eq.~\eqref{eq:bornmarkov3}
\begin{equation}
\label{eq:substperlind}
\begin{aligned}
\Gamma_{k,k'}^\alpha(\omega_j)&\rightarrow\Gamma_{k,k'}^\alpha(\omega_j,\omega_{j'})\equiv \sqrt{\Gamma_{k,k'}^\alpha(\omega_j)\Gamma_{k,k'}^\alpha(\omega_{j'})},\\
\Delta_{k,k'}^\alpha(\omega_j)&\rightarrow\Delta_{k,k'}^\alpha(\omega_j,\omega_{j'})\equiv\frac{\Delta_{k,k'}^\alpha(\omega_j)+\Delta_{k,k'}^\alpha(\omega_{j'})}{2}.
\end{aligned}
\end{equation}
This results in the PERLind master equation which is of GKLS form
\begin{equation}
\label{eq:perlind}
\begin{aligned}
\partial_t\hat{\rho}_{\rm S}=-i[\hat{H}_{\rm S},\hat{\rho}_{\rm S}]+\sum_{\alpha}\sum_{k,k'}\sum_{j,j'}&\left\{\Gamma_{k,k'}^\alpha(\omega_j,\omega_{j'})\left[\hat{S}^{j}_{\alpha,k'}\hat{\rho}_{\rm S}\left(\hat{S}^{j'}_{\alpha,k}\right)^\dagger-\frac{1}{2}\left\{\left(\hat{S}^{j'}_{\alpha,k}\right)^\dagger\hat{S}^j_{\alpha,k'},\hat{\rho}_{\rm S}\right\}\right]\right.\\&\left.
-i\Delta^\alpha_{k,k'}(\omega_j,\omega_{j'})\left[\left(\hat{S}^{j'}_{\alpha,k}\right)^\dagger\hat{S}^j_{\alpha,k'},\hat{\rho}_{\rm S}\right]\right\}.
\end{aligned}
\end{equation}
This approach interpolates between the second Markov approximation and the secular approximation and is expected to be applicable whenever the Born-Markov approximations are justified. While the second law is not guaranteed to hold, any violations are of higher order in the system-bath coupling and can therefore be neglected due to the perturbative nature of the Born-Markov approximations \cite{kirsanskas:2018,ptaszynski:2019}.

\section{Single-component systems}
In this section, we consider systems which satisfy Eqs.~\eqref{eq:sjump} and \eqref{eq:restb} with a single frequency $\omega$. For lack of a better word, we call these systems single-component systems because we find that these conditions are fulfilled in physical scenarios for a single quantum dot, qubit, or harmonic oscillator.

\subsection{Single bath - thermalization}
We first consider a spinless, single-level quantum dot tunnel-coupled to a single fermionic bath. The Hamiltonian of system and bath is then given by
\begin{equation}
\label{eq:hamtotsingd}
\hat{H}_{\rm tot}=\hat{H}_{\rm S}+\hat{H}_{\rm B}+\hat{H}_{\rm BS},
\end{equation}
with 
\begin{equation}
\label{eq:hamtotsingd2}
\hat{H}_{\rm S}=\varepsilon_{\rm d}\hat{d}^\dagger\hat{d},\hspace{1.5cm}\hat{H}_{\rm B}=\sum_q\varepsilon_q\hat{c}_q^\dagger\hat{c}_q,\hspace{1.5cm}\hat{H}_{\rm BS}=\hat{d}\sum_qg_q\hat{c}_q^\dagger-\hat{d}^\dagger\sum_qg_q^*\hat{c}_q.
\end{equation}
Since
\begin{equation}
[\hat{d},\hat{H}_{\rm S}]=\varepsilon_{\rm d}\hat{d},
\end{equation}
We can use Eq.~\eqref{eq:mastersingcomp} with $\hat{S}_0=\hat{d}$, $\hat{S}_1=\hat{d}^\dagger$, and $\hat{B}_0=\sum_qg_q\hat{c}^\dagger$, $\hat{B}_1=\sum_qg_q^*\hat{c}_q$. In this case, the bath correlation functions read
\begin{equation}
\label{eq:bathcorrferm}
\begin{aligned}
C_{1,1}(s)&=\sum_qe^{i\varepsilon_qs}|g_q|^2n_{\rm F}(\varepsilon_q)=\int_{-\infty}^\infty d\omega e^{i\omega s}\rho(\omega)n_{\rm F}(\omega),\\
C_{0,0}(s)&=\sum_qe^{-i\varepsilon_qs}|g_q|^2[1-n_{\rm F}(\varepsilon_q)]=\int_{-\infty}^\infty d\omega e^{-i\omega s}\rho(\omega)[1-n_{\rm F}(\omega)],
\end{aligned}
\end{equation}
where we introduced the spectral density
\begin{equation}
\label{eq:specdens}
\rho(\omega)=\sum_q|g_q|^2\delta(\varepsilon_q-\omega),
\end{equation}
which can be treated as a continuous function in the limit of a large bath with many modes. 

From Eq.~\eqref{eq:gammas} and \eqref{eq:deltas}, we find
\begin{equation}
\begin{aligned}
&\gamma_1=\kappa n_{\rm F}(\varepsilon_{\rm d}),\hspace{3.75cm}
\gamma_0=\kappa[1-n_{\rm F}(\varepsilon_{\rm d})],\\
&\Delta_1 = -P\int_{-\infty}^{\infty}\frac{\rho(\omega)}{\varepsilon_{\rm d}-\omega}n_{\rm F}(\omega),\hspace{1.5cm}\Delta_0 = P\int_{-\infty}^{\infty}\frac{\rho(\omega)}{\varepsilon_{\rm d}-\omega}[1-n_{\rm F}(\omega)]
\end{aligned}
\end{equation}
where we introduced $\kappa=2\pi\rho(\varepsilon_{\rm d})$, $P$ denotes the Cauchy principal value, and we made use of
\begin{equation}
\label{eq:intsing}
\lim\limits_{t\rightarrow\infty}\int_{0}^{t}ds e^{i\Omega s}=\pi\delta(\Omega)+iP\left(\frac{1}{\Omega}\right).
\end{equation}
Finally, inserting all quantities in Eq.~\eqref{eq:mastersingcomp} results in the Markovian master equation
\begin{equation}
\label{eq:masterdottherm}
\partial_t\hat{\rho}_{\rm S} = -i[\bar{\varepsilon}_{\rm d}\hat{d}^\dagger\hat{d},\hat{\rho}_{\rm S}]+\kappa[1-n_{\rm F}(\varepsilon_{\rm d})]\mathcal{D}[\hat{d}]\hat{\rho}_{\rm S}+\kappa n_{\rm F}(\varepsilon_{\rm d})\mathcal{D}[\hat{d}^\dagger]\hat{\rho}_{\rm S},
\end{equation}
where the renormalized dot energy reads
\begin{equation}
\label{eq:varepsbar}
\bar{\varepsilon}_{\rm d} = \varepsilon_{\rm d}+P\int_{-\infty}^{\infty}d\omega\frac{\rho(\omega)}{\varepsilon_{\rm d}-\omega},
\end{equation}
and we introduced the superoperators
\begin{equation}
\label{eq:superlindd}
\mathcal{D}[\hat{A}]\hat{\rho}=\hat{A}\hat{\rho}\hat{A}^\dagger-\frac{1}{2}\{\hat{A}^\dagger\hat{A},\hat{\rho}\}.
\end{equation}
The bath thus has two effects. Through the dissipative part of the master equations, it describes electrons entering ($\mathcal{D}[\hat{d}^\dagger]$) and leaving ($\mathcal{D}[\hat{d}]$) the quantum dot. In addition, the energy level of the quantum dot is renormalized.

As discussed above, the Markovian approximation is justified if the bath correlation functions decay on a time scale which is much faster than the time over which $\tilde{\rho}_{\rm S}$ varies. In the limiting case where both $\rho(\omega)$ as well as $n_{\rm F}(\omega)$ are independent of $\omega$, the bath correlation functions become proportional to a Dirac delta function and the bath becomes truly \textit{memoryless} (i.e., $\tau_{\rm B}\rightarrow0$). In practice, it is sufficient for $\tau_{\rm B}$ to be much shorter than any relevant time-scale of the system. In energy, this translates to the condition that the functions $\rho(\omega)$ as well as $n_{\rm F}(\omega)$ are flat around the relevant energies of the system. For the present system $\tilde{\rho}_{\rm S}$ changes on the time-scale $1/\kappa$. The Markov approximation is then valid as long as $\kappa\tau_{\rm B}\ll 1$. In energy space, this requires $\rho(\omega)$ as well as $n_{\rm F}(\omega)$ to be approximately constant in the interval $\omega\in[\varepsilon_{\rm d}-\kappa,\varepsilon_{\rm d}+\kappa]$.
 At low temperatures, the Fermi-Dirac distribution becomes a step function. Therefore, the Markovian approximation is not justified at low temperatures if the dot level $\varepsilon_{\rm d}$ is close to the chemical potential.

\subsubsection{Solving the master equation} To solve the master equation, we write the density matrix of the system as
\begin{equation}
\hat{\rho}_S=p_0|0\rangle\langle 0|+p_1|1\rangle\langle 1|,
\end{equation}
where we used the fact that we cannot have a superposition of states with a different number of electrons in the system due to particle number conservation. From Eq.~\eqref{eq:masterdottherm}, we then find
\begin{equation}
\label{eq:dtp0dot}
\partial_t p_1 = -\kappa[1-n_{\rm F}(\varepsilon_{\rm d})]p_1+\kappa n_{\rm F}(\varepsilon_{\rm d})p_0 = -\kappa[p_1-n_{\rm F}(\varepsilon_{\rm d})].
\end{equation}
This equation shows that a full dot is emptied with rate $\kappa[1-n_{\rm F}(\varepsilon_{\rm d})]$ whereas an empty dot is filled with rate $\kappa n_{\rm F}(\varepsilon_{\rm d})$. The solution of the differential equation reads
\begin{equation}
\label{eq:solutiondottherm}
p_1(t)=p_1(0)e^{-\kappa t}+n_{\rm F}(\varepsilon_{\rm d})(1-e^{-\kappa t}).
\end{equation}
The occupation probability thus exponentially goes toward the equilibrium value $n_{\rm F}(\varepsilon_{\rm d})$. The time-scale with which this happens is given by $1/\kappa$. In equilibrium, the system is described by the thermal state as expected and we find
\begin{equation}
\label{eq:longtermdottherm}
\lim\limits_{t\rightarrow\infty}\hat{\rho}_{\rm S} = \frac{e^{-\beta(\varepsilon_{\rm d}-\mu)\hat{d}^\dagger\hat{d}}}{{\rm Tr}\{e^{-\beta(\varepsilon_{\rm d}-\mu)\hat{d}^\dagger\hat{d}}\}}.
\end{equation}

\subsubsection{Energy flows and the first law} In addition to the state of the quantum dot, we are interested in the energy flow between the dot and the bath. In Sec.~\ref{sec:entropy}, we have seen that the power supplied by the bath can be expressed as the change in particle number induced by the bath. We thus consider
\begin{equation}
\label{eq:numchdot}
\partial_t\langle \hat{d}^\dagger\hat{d}\rangle = {\rm Tr}\{\hat{d}^\dagger\hat{d}\partial_t\hat{\rho}_{\rm S}\}=\kappa[1-n_{\rm F}(\varepsilon_{\rm d})]{\rm Tr}\{\hat{d}^\dagger\hat{d}\mathcal{D}[\hat{d}]\hat{\rho}_{\rm S}\}+\kappa n_{\rm F}(\varepsilon_{\rm d}){\rm Tr}\{\hat{d}^\dagger\hat{d}\mathcal{D}[\hat{d}^\dagger]\hat{\rho}_{\rm S}\},
\end{equation}
where we used Eq.~\eqref{eq:masterdottherm} as well as ${\rm Tr}\{\hat{A}[\hat{A},\hat{B}]\}=0$. From Eq.~\eqref{eq:numchdot}, we see that all changes in the particle number are mediated by the bath. This is a consequence of the fact that the Hamiltonian conserves the number of particles and that there is only a single bath. Since $\langle \hat{d}^\dagger\hat{d}\rangle=p_1$, Eq.~\eqref{eq:numchdot} is equivalent to Eq.~\eqref{eq:dtp0dot} and we find for the power that flows into the bath
\begin{equation}
\label{eq:powerbathdot}
P_{\rm B}(t) = -\mu\partial_t\langle \hat{d}^\dagger\hat{d}\rangle= \mu\kappa e^{-\kappa t}[p_1(0)-n_{\rm F}(\varepsilon_{\rm d})].
\end{equation}
Similarly, we find for the heat flow
\begin{equation}
\label{eq:heatbathdot}
J_{\rm B}(t) = -(\varepsilon_{\rm d}-\mu)\partial_t\langle \hat{d}^\dagger\hat{d}\rangle= (\varepsilon_{\rm d}-\mu)\kappa e^{-\kappa t}[p_1(0)-n_{\rm F}(\varepsilon_{\rm d})].
\end{equation}
We thus find that if the dot starts out in a non-equilibrium state, there is an exponentially decreasing energy flow which can be divided into power and heat. The first law is obeyed by construction [cf.~Eq.~\eqref{eq:firstlaw}]
\begin{equation}
\label{eq:firstlawdot}
\partial_t\langle \hat{H}_{\rm S}\rangle = \varepsilon_{\rm d}\partial_t\langle \hat{d}^\dagger\hat{d}\rangle=-P_{\rm B}-J_{\rm B}.
\end{equation}

\subsubsection{Entropy and the second law} For the second law, we need to consider the entropy of the quantum dot given by
\begin{equation}
\label{eq:vndot}
S_{\rm vN}[\hat{\rho}_{\rm S}]=-p_0\ln p_0 -p_1\ln p_1,\hspace{1.5cm}\partial_tS_{\rm vN}[\hat{\rho}_{\rm S}] = -\dot{p}_1\ln\frac{p_1}{1-p_1},
\end{equation}
where the dot denotes the time-derivative and we used $p_0+p_1=1$ to compute the derivative. The entropy production rate given in Eq.~\eqref{eq:entprod} can then be expressed as
\begin{equation}
\label{eq:entproddot}
\Sigma = \partial_tS_{\rm vN}[\hat{\rho}_{\rm S}]+\frac{J_{\rm B}}{k_BT}=\frac{J_{\rm B}}{(\varepsilon_{\rm d}-\mu)}\left[\beta(\varepsilon_{\rm d}-\mu)+\ln\frac{p_1}{1-p_1}\right],
\end{equation}
where we used $-\dot{p_1}=J_{\rm B}/(\varepsilon_{\rm d}-\mu)$ which follows from Eq.~\eqref{eq:heatbathdot}. Using the equality
\begin{equation}
\label{eq:fermisexp}
e^{\beta(\varepsilon_{\rm d}-\mu)}=\frac{1-n_{\rm F}(\varepsilon_{\rm d})}{n_{\rm F}(\varepsilon_{\rm d})},
\end{equation}
we find
\begin{equation}
\label{eq:entproddot2}
\begin{aligned}
\Sigma =&\frac{J_{\rm B}}{(\varepsilon_{\rm d}-\mu)}\ln\left(\frac{p_1[1-n_{\rm F}(\varepsilon_{\rm d})]}{[1-p_1]n_{\rm F}(\varepsilon_{\rm d})}\right)\\=&\kappa\Big(p_1[1-n_{\rm F}(\varepsilon_{\rm d})]-[1-p_1]n_{\rm F}(\varepsilon_{\rm d})\Big)\ln\left(\frac{p_1[1-n_{\rm F}(\varepsilon_{\rm d})]}{[1-p_1]n_{\rm F}(\varepsilon_{\rm d})}\right)\geq 0,
\end{aligned}
\end{equation}
where positivity of $\Sigma$ can be shown by writing it in the form $(x-y)(\ln x-\ln y)\geq 0$ which is ensured to be positive because the logarithm is a monotonously increasing function. Using the solution in Eq.~\eqref{eq:solutiondottherm}, we can explicitly write the entropy production rate as
\begin{equation}
\Sigma = \kappa e^{-\kappa t}\delta_0\ln\frac{(e^{-\kappa t}\delta_0+n_{\rm F})(1-n_{\rm F})}{(1-n_{\rm F}-e^{-\kappa t}\delta_0)n_{\rm F}},
\end{equation}
where we introduced $\delta_0 = p_1(0)-n_{\rm F}$ and we suppressed the argument of the Fermi-Dirac distribution for ease of notation.

{\small 
	\paragraph{Exercises}
	
	\paragraph{2.4 Markovian master equation for a qubit in an electromagnetic field (2pt)}
	
	Derive the (Born-Markov) Markovian master equation for a qubit coupled to a bosonic reservoir with $\mu=0$ described by the total Hamiltonian
	\begin{equation}
	\label{eq:hamtotquex}
	\hat{H}_{\rm tot}=\hat{H}_{\rm S}+\hat{H}_{\rm B}+\hat{H}_{\rm BS},
	\end{equation}
	with 
	\begin{equation}
	\label{eq:hamtotquex2}
	\hat{H}_{\rm S}=\frac{\varepsilon_{\rm S}}{2}\hat{\sigma}_{\rm z},\hspace{1.5cm}\hat{H}_{\rm B}=\sum_q\varepsilon_{\rm q}\hat{a}_q^\dagger\hat{a}_q,\hspace{1.5cm}\hat{H}_{\rm BS}=\sum_q\left(g_q\hat{\sigma}\hat{a}_q^\dagger+g_q^*\hat{\sigma}^\dagger \hat{a}_q\right),
	\end{equation}
	where $\hat{\sigma}_{\rm z}$ is the Pauli z-matrix, $\hat{a}_q$ is a bosonic annihilation operator, and we introduced the raising and lowering operators for the qubit
	\begin{equation}
	\hat{\sigma}^\dagger = |1\rangle\langle 0|,\hspace{1.5cm}\hat{\sigma}=|0\rangle\langle 1|.
	\end{equation}
	Compare to Eq.~\eqref{eq:masterdottherm}.
	
}

\subsection{Two baths - heat engine}
\label{sec:qdhe}
In this section, we consider a simplified version of the heat engine that was implemented experimentally in Ref.~\cite{josefsson:2018}. In contrast to a quantum dot coupled to a single bath, where the only thing that happens is thermalization, we will find heat flows in the steady state and we will see how heat can be converted into work and how work can be used to refrigerate. The system we consider is a spinless, single-level quantum dot tunnel-coupled to two heat baths
\begin{equation}
\label{eq:hamtotdh}
\hat{H}_{\rm tot}=\hat{H}_{\rm S}+\hat{H}_{\rm C}+\hat{H}_{\rm H}+\hat{H}_{\rm CS}+\hat{H}_{\rm HS},
\end{equation}
with
\begin{equation}
\label{eq:hamtotdh2}
\hat{H}_{\rm S}=\varepsilon_{\rm d}\hat{d}^\dagger\hat{d},\hspace{1cm}\hat{H}_{\rm \alpha}=\sum_q\varepsilon_{\alpha,q}\hat{c}_{\alpha,q}^\dagger\hat{c}_{\alpha,q},\hspace{1cm}\hat{H}_{\alpha{\rm S}}=\hat{d}\sum_qg_{\alpha,q}\hat{c}_{\alpha,q}^\dagger-\hat{d}^\dagger\sum_qg_{\alpha,q}^*\hat{c}_{\alpha,q},
\end{equation}
where $\alpha={\rm C, H}$ labels the baths according to their temperatures $T_{\rm C}\leq T_{\rm H}$.
Just as for the quantum dot coupled to a single bath, the conditions to use Eq.~\eqref{eq:mastersingcomp} are fulfilled. Since the terms in the master equation induced by different baths are additive, we find
\begin{equation}
\label{eq:masterdh}
\partial_t\hat{\rho}_{\rm S} = -i[\hat{H}_{\rm S},\hat{\rho}_{\rm S}]+\mathcal{L}_{\rm C}\hat{\rho}_{\rm S}+\mathcal{L}_{\rm H}\hat{\rho}_{\rm S},
\end{equation}
with
\begin{equation}
\label{eq:bathlind}
\mathcal{L}_\alpha\hat{\rho}=\kappa_\alpha[1-n_{\rm F}^\alpha(\varepsilon_{\rm d})]\mathcal{D}[\hat{d}]\hat{\rho}+\kappa_\alpha n^\alpha_{\rm F}(\varepsilon_{\rm d})\mathcal{D}[\hat{d}^\dagger]\hat{\rho},
\end{equation}
where $n^\alpha_{\rm F}$ is the Fermi-Dirac distribution with temperature $T_\alpha$ and chemical potential $\mu_\alpha$. Here we neglected the renormalization of $\varepsilon_{\rm d}$ which is given by a straightforward generalization of Eq.~\eqref{eq:varepsbar}.

\subsubsection{Solving the master equation} 
The master equation can easily be solved by considering
\begin{equation}
\partial t p_1 = {\rm Tr}\{\hat{d}^\dagger\hat{d}\partial_t\hat{\rho}_{\rm S}\}=-\sum_{\alpha={\rm C,H}}\kappa_\alpha\left\{[1-n_{\rm F}^\alpha(\varepsilon_{\rm d})]p_1-n_{\rm F}^\alpha(\varepsilon_{\rm d})p_0\right\} = -\gamma(p_1-\bar{n}),
\end{equation}
where
\begin{equation}
\label{eq:gamman}
\gamma = \kappa_{\rm C}+\kappa_{\rm H},\hspace{2cm}\bar{n}=\frac{\kappa_{\rm C}n_{\rm F}^{\rm C}(\varepsilon_{\rm d})+\kappa_{\rm H}n_{\rm F}^{\rm H}(\varepsilon_{\rm d})}{\kappa_{\rm C}+\kappa_{\rm H}}.
\end{equation}
Comparing to Eq.~\eqref{eq:dtp0dot}, we find that the quantum dot behaves just like a quantum dot coupled to a single heat bath with coupling strength $\gamma$ and mean occupation $\bar{n}$. The solution thus reads
\begin{equation}
\label{eq:solutiondh}
p_1(t)=p_1(0)e^{-\gamma t}+\bar{n}(1-e^{-\gamma t}).
\end{equation}

\subsubsection{Energy flows and the first law} 
Since there are now two baths, the number of electrons on the dot can change due to either bath. Indeed, we can write
\begin{equation}
\label{eq:numchangbath}
\partial_t\langle \hat{d}^\dagger\hat{d}\rangle = {\rm Tr}\{\hat{d}^\dagger\hat{d}\mathcal{L}_{\rm C}\hat{\rho}_{\rm S}\}+{\rm Tr}\{\hat{d}^\dagger\hat{d}\mathcal{L}_{\rm H}\hat{\rho}_{\rm S}\}.
\end{equation}
We thus find
\begin{equation}
\label{eq:powerheateachbath}
\begin{aligned}
P_\alpha =& -\mu_\alpha{\rm Tr}\{\hat{d}^\dagger\hat{d}\mathcal{L}_\alpha\hat{\rho}_{\rm S}\},\\
J_\alpha =&-(\varepsilon_{\rm d}-\mu_\alpha){\rm Tr}\{\hat{d}^\dagger\hat{d}\mathcal{L}_\alpha\hat{\rho}_{\rm S}\},
\end{aligned}
\end{equation}
where the first law of thermodynamics is again obeyed by construction.
Explicitly, we find
\begin{equation}
\label{eq:poweralphadh}
P_\alpha = \mu_\alpha\kappa_\alpha e^{-\gamma t}[p_1(0)-\bar{n}]+\mu_\alpha\kappa_\alpha[\bar{n}-n_{\rm F}^\alpha(\varepsilon_{\rm d})],
\end{equation}
and similarly for the heat currents upon replacing $\mu_\alpha\rightarrow\varepsilon_{\rm d}-\mu_\alpha$. Just as for a single bath, there is a transient term in the power which decreases exponentially in time. In contrast to the single bath case, there is now also a time-independent term which remains in steady state. 
%This term is of equal magnitude and opposite signs for the two baths since
%\begin{equation}
%\kappa_{\rm C}[\bar{n}-n_{\rm F}^{\rm C}(\varepsilon_{\rm d})] = \frac{\kappa_{\rm C}\kappa_{\rm H}}{\kappa_{\rm C}+\kappa_{\rm H}}[n_{\rm F}^{\rm H}(\varepsilon_{\rm d})-n_{\rm F}^{\rm C}(\varepsilon_{\rm d})]=-\kappa_{\rm H}[\bar{n}-n_{\rm F}^{\rm H}(\varepsilon_{\rm d})].
%\end{equation}

\subsubsection{Steady state} 
In the steady state, the observables of the system do not change. We can use this fact to draw a number of conclusions without using the explicit solutions for the power and the heat currents. In particular, since the left-hand side of Eq.~\eqref{eq:numchangbath} vanishes, we find
\begin{equation}
\label{eq:steadych}
{\rm Tr}\{\hat{d}^\dagger\hat{d}\mathcal{L}_{\rm C}\hat{\rho}_{\rm S}\}=-{\rm Tr}\{\hat{d}^\dagger\hat{d}\mathcal{L}_{\rm H}\hat{\rho}_{\rm S}\}.
\end{equation}
From this, using Eqs.~\eqref{eq:powerheateachbath}, follows
\begin{equation}
\label{eq:firstlawss2t}
P=P_{\rm C}+P_{\rm H} = -(J_{\rm C}+J_{\rm H}),
\end{equation}
which is nothing but the first law,
as well as
\begin{equation}
\eta = \frac{P}{-J_{\rm H}}=\frac{\mu_{\rm C}-\mu_{\rm H}}{\varepsilon_{\rm d}-\mu_{\rm H}}=1-\frac{\varepsilon_{\rm d}-\mu_{\rm C}}{\varepsilon_{\rm d}-\mu_{\rm H}},
\end{equation}
where we introduced the efficiency $\eta$ which is given by the ratio between the power (the output of the heat engine) and the heat current from the hot bath (the input of the heat engine). Using the explicit solution for the power in Eq.~\eqref{eq:poweralphadh}, we find
\begin{equation}
\label{eq:powerssdh}
P=\frac{\kappa_{\rm C}\kappa_{\rm H}}{\kappa_{\rm C}+\kappa_{\rm H}}(\mu_{\rm C}-\mu_{\rm H})[n_{\rm F}^{\rm H}(\varepsilon_{\rm d})-n_{\rm F}^{\rm C}(\varepsilon_{\rm d})].
\end{equation}

\begin{figure}
	\centering
	\includegraphics[width=\textwidth]{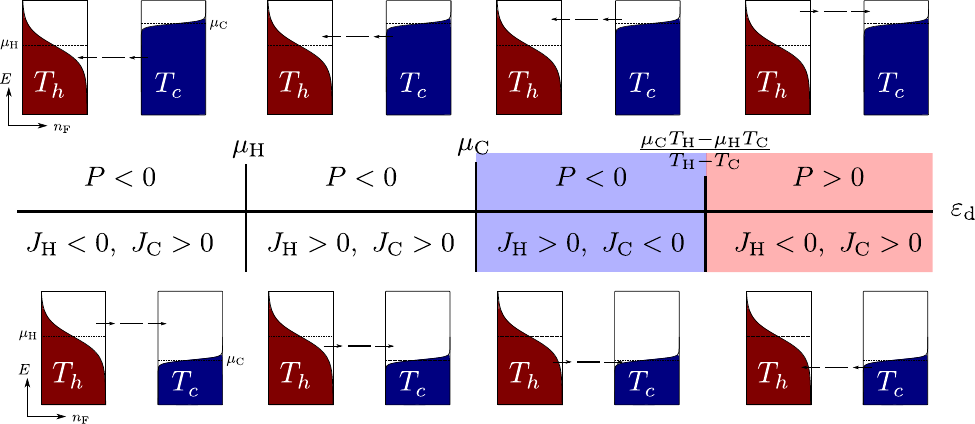}
	\caption{Different regimes determined by the signs of $P$, $J_{\rm C}$, and $J_{\rm H}$. For $\mu_{\rm C}\geq\mu_{\rm H}$, $\varepsilon_{\rm d}$ increases from left to right (illustrated by the cartoons in the upper row). For $\mu_{\rm C}\leq\mu_{\rm H}$, we find the exact same regimes but $\varepsilon_{\rm d}$ decreases from left to right (illustrated by the cartoons in the lower row). In the regime to the far left, heat flows out of the hot bath and into the cold bath and power is dissipated. If $\varepsilon_{\rm d}$ lies in between the chemical potentials, both baths are heated up by the dissipated power. Refrigeration is obtained in the blue shaded regime, where heat is extracted from the cold bath and dumped into the hot bath. Finally, in the red shaded regime the quantum dot acts as a heat engine where a heat flow from hot to cold drives a charge flow against the external voltage bias.}
	\label{fig:qd_regimes}
\end{figure}

\begin{figure}
	\centering
	\includegraphics[width=\textwidth]{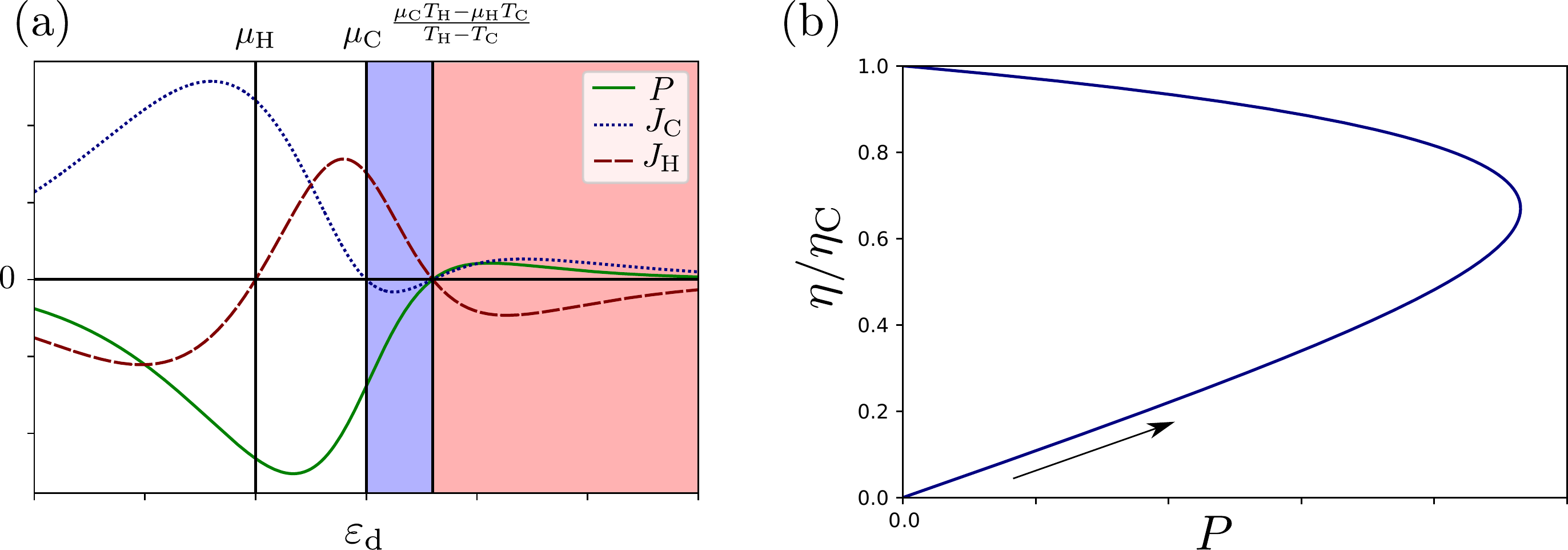}
	\caption{Performance of the quantum dot heat engine. (a) Power and heat currents as a function of $\varepsilon_{\rm d}$. The regimes illustrated in Fig.~\ref{fig:qd_regimes} can clearly be identified. (b) ``Lasso" diagram. Along the curve, the voltage bias is increased in the direction of the arrow. For $\mu_{\rm C}=\mu_{\rm H}$, both the power as well as the efficiency vanishes. Increasing $\mu_{\rm C}$ then results in a finite power and efficiency, until the power vanishes again at the stopping voltage, where the efficiency reaches $\eta_{\rm C}$. Such plots are called Lasso diagrams because the curve usually goes back to the origin at the stopping voltage. Parameters (unit-less): $\varepsilon_{\rm d}=2$, $T_{\rm C}=0.3$, $T_{\rm H}=0.8$, $\mu_{\rm C}=1$, $\mu_{\rm H}=0$, $\kappa_{\rm C}=\kappa_{\rm H}=1$.}
	\label{fig:qd_ed}
\end{figure}

Let us now consider under which conditions the system acts as a heat engine, i.e., heat from the hot bath is converted into power. From Eq.~\eqref{eq:powerssdh}, we can identify different regimes depending on the signs of $P$, $J_{\rm C}$, and $J_{\rm H}$. These regimes are illustrate in Figs.~\ref{fig:qd_regimes} and \ref{fig:qd_ed}\,(a). For $\mu_{\rm C}\geq\mu_{\rm H}$ ($\mu_{\rm C}\leq\mu_{\rm H}$), we find that the quantum dot acts as a heat engine for large positive (negative) values of $\varepsilon_{\rm d}$. In both cases, we find from Eq.~\eqref{eq:powerssdh} that power is positive as long as
\begin{equation}
\frac{\varepsilon_{\rm d}-\mu_{\rm C}}{\varepsilon_{\rm d}-\mu_{\rm H}}\geq\frac{T_{\rm C}}{T_{\rm H}}\hspace{.5cm}\Rightarrow\hspace{.5cm}\eta\leq 1-\frac{T_{\rm C}}{T_{\rm H}}=\eta_{\rm C}.
\end{equation}
We thus find that the efficiency is bounded from above by the Carnot efficiency as long as the power output is non-negative.

\subsubsection{Entropy and the second law} 
We first make some general statements about the second law in a two-terminal setup. The entropy production rate is given by
\begin{equation}
\label{eq:secondlawss2}
\Sigma = \partial_tS_{\rm vN}[\hat{\rho}_{\rm S}]+\frac{J_{\rm C}}{k_BT_{\rm C}}+\frac{J_{\rm H}}{k_BT_{\rm H}}\geq 0.
\end{equation}
In the steady state, the first term vanishes and we immediately find that at least one of the heat currents has to be positive. This implies that it is impossible to cool down all baths at the same time (in the steady state). Furthermore, for equal temperatures ($T_{\rm C}=T_{\rm H}$), we find $P=-(J_{\rm C}+J_{\rm H})\leq 0$ which implies that it is not possible to convert heat into work with baths at a single temperature. This is known as the Kelvin-Planck statement of the second law. Finally, we can use the first law in Eq.~\eqref{eq:firstlawss2t} to eliminate $J_{\rm C}$, resulting in
\begin{equation}
\Sigma = \frac{P}{k_{\rm B}T_{\rm C}}\frac{\eta_{\rm C}-\eta}{\eta}\hspace{.5cm} \Rightarrow\hspace{.5cm} 0\leq\eta\leq\eta_{\rm C},
\end{equation}
where the last inequality holds for $P\geq 0$. The fact that the efficiency is upper bounded by the Carnot efficiency is thus a direct consequence of the second law.

The interplay between power and efficiency is illustrated in Fig.~\ref{fig:qd_ed}. We find that at maximum power, the efficiency reaches above $60\%$ of the Carnot efficiency. Similar values where found experimentally in Ref.~\cite{josefsson:2018}. We note that at the stopping voltage, the Carnot efficiency is obtained. This is a consequence of the fact that there is only a single energy at which transport happens. At the stopping voltage, all transport is blocked implying that both the charge as well as the heat currents vanish. This implies that there is no dissipation ($\Sigma=0$) and the efficiency takes on the Carnot value (see also Ref.~\cite{brunner:2012}). If multiple energies are involved, transport happens at all energies even at the stopping voltage, where the charge currents sum up to zero. There can however still be a net heat current which then results in a finite entropy production and a vanishing efficiency. In this case, Fig.~\ref{fig:qd_ed}\,(b) takes on the shape of a lasso. In the experiment of Ref.~\cite{josefsson:2018}, imperfections (that may be unavoidable) spoil the perfect energy filtering aspect of the quantum dot and Carnot efficiency cannot be reached.

In our system, the entropy of the quantum dot is given in Eq.~\eqref{eq:vndot}. Using
\begin{equation}
\partial_t p_1 = -\frac{J_{\rm C}}{\varepsilon_{\rm d}-\mu_{\rm C}}-\frac{J_{\rm H}}{\varepsilon_{\rm d}-\mu_{\rm H}},
\end{equation}
we can write the entropy production rate as
\begin{equation}
\label{eq:entdh}
\Sigma = \sum_{\alpha={\rm C,H}}\kappa_\alpha\Big(p_1[1-n^\alpha_{\rm F}(\varepsilon_{\rm d})]-[1-p_1]n^\alpha_{\rm F}(\varepsilon_{\rm d})\Big)\ln\left(\frac{p_1[1-n^\alpha_{\rm F}(\varepsilon_{\rm d})]}{[1-p_1]n^\alpha_{\rm F}(\varepsilon_{\rm d})}\right)\geq 0,
\end{equation}
which is positive since each term in the sum is positive in complete analogy to Eq.~\eqref{eq:entproddot2}. In the steady state, we find
\begin{equation}
\label{eq:entdhss}
\Sigma = \frac{\kappa_{\rm C}\kappa_{\rm H}}{\kappa_{\rm C}+\kappa_{\rm H}}[n_{\rm F}^{\rm H}(\varepsilon_{\rm d})-n_{\rm F}^{\rm C}(\varepsilon_{\rm d})]\left[\beta_{\rm C}(\varepsilon_{\rm d}-\mu_{\rm C})-\beta_{\rm H}(\varepsilon_{\rm d}-\mu_{\rm H})\right]\geq 0,
\end{equation}
which vanishes at the Carnot point, where $n_{\rm F}^{\rm H}(\varepsilon_{\rm d})=n_{\rm F}^{\rm C}(\varepsilon_{\rm d})$, $\eta=\eta_{\rm C}$, and both the power as well as the heat currents vanish. We stress that while an equilibrium situation (i.e., $T_{\rm C}=T_{\rm H}$ and $\mu_{\rm C}=\mu_{\rm H}$) ensures $n_{\rm F}^{\rm H}(\varepsilon_{\rm d})=n_{\rm F}^{\rm C}(\varepsilon_{\rm d})$, the Carnot point can also be reached out of equilibrium. Indeed, only for $T_{\rm C}<T_{\rm H}$ is the Carnot efficiency finite.

\subsubsection{Refrigeration} 
As discussed above, the quantum dot can also act as a refrigerator in the regime where $P<0$, $J_{\rm H}> 0$, and $J_{\rm C}< 0$. In this case, electrical power is used to reverse the natural flow of heat, resulting in a heat flow out of the cold bath and into the hot bath. The efficiency of this process is usually characterized by the coefficient of performance (COP)
\begin{equation}
\label{eq:cop}
\eta^{\rm COP} = \frac{-J_{\rm C}}{-P},
\end{equation}
where we left the minus signs to stress that this performance quantifier is relevant in the regime where both $P$ as well as $J_{\rm C}$ are negative under our sign convention. We can use the first law in Eq.~\eqref{eq:firstlawss2t} to eliminate $J_{\rm H}$ and write the entropy production rate as
\begin{equation}
\Sigma = \frac{-J_{\rm C}}{k_{\rm B}T_{\rm H}}\frac{\eta_{\rm C}^{\rm COP}-\eta^{\rm COP}}{\eta_{\rm C}^{\rm COP}\eta^{\rm COP}}\hspace{.5cm}\Rightarrow\hspace{.5cm}0\leq\eta^{\rm COP}\leq \eta_{\rm C}^{\rm COP},
\end{equation}
where the second inequality holds for $J_{\rm C}\leq 0$ and we introduced the Carnot value for the COP
\begin{equation}
\label{eq:copcarnot}
\eta_{\rm C}^{\rm COP} = \frac{T_{\rm C}}{T_{\rm H}-T_{\rm C}}.
\end{equation}
We note that as $T_{\rm C}\rightarrow T_{\rm H}$, $\eta_{\rm C}^{\rm COP}$ diverges. This reflects the fact that in principle, it is possible to move heat in between two baths with equal temperature without investing any work.

In our system, we find from Eqs.~\eqref{eq:powerheateachbath} and \eqref{eq:steadych}
\begin{equation}
\eta^{\rm COP} = \frac{\mu_{\rm C}-\varepsilon_{\rm d}}{\mu_{\rm C}-\mu_{\rm H}}.
\end{equation}
From this, we find that $\eta^{\rm COP}$ vanishes when $\varepsilon_{\rm d}=\mu_{\rm C}$ and takes on the Carnot value at $\varepsilon_{\rm d}=\frac{\mu_{\rm C}T_{\rm H}-\mu_{\rm H}T_{\rm C}}{T_{\rm H}-T_{\rm C}}$, which is exactly the point where the regime of the refrigerator meets the regime of the heat engine, see Fig.~\ref{fig:qd_regimes}. Interestingly, both the COP as well as the efficiency reach their maximum value at this point, where no transport takes place. The Carnot point is often called the point of \textit{reversibility}. At this point nothing happens but it can be seen as the limit of converting heat into work infinitely slowly and without wasting any energy (thus, reversibly). Equivalently, taking the limit from the other side, it can be seen as the limit of cooling the cold bath reversibly.

{\small 
	\paragraph{Exercises}
	
	\paragraph{3.1 Heat engine with external power (4pt)}
	Consider the Markovian master equation
	\begin{equation}
	\label{eq:meheex}
	\partial_t\hat{\rho}_{\rm S} = -i[\hat{H}_{\rm S}(t),\hat{\rho}_{\rm S}]+\mathcal{L}_{\rm C}\hat{\rho}_{\rm S}+\mathcal{L}_{\rm H}\hat{\rho}_{\rm S},
	\end{equation}
	with the time-dependent system Hamiltonian
	\begin{equation}
\label{eq:mehehs}
	\hat{H}_{\rm S}(t) = \Omega_{\rm C}\hat{a}_{\rm C}^\dagger\hat{a}_{\rm C}+\Omega_{\rm H}\hat{a}_{\rm H}^\dagger\hat{a}_{\rm H}+g\left(\hat{a}_{\rm C}^\dagger\hat{a}_{\rm H}e^{i(\Omega_{\rm H}-\Omega_{\rm C})t}+\hat{a}_{\rm H}^\dagger\hat{a}_{\rm C}e^{-i(\Omega_{\rm H}-\Omega_{\rm C})t}\right),
	\end{equation}
	and the dissipators
	\begin{equation}
	\label{eq:mehediss}
	\mathcal{L}_\alpha\hat{\rho}=\kappa_\alpha[1+n_{\rm B}^\alpha(\Omega_\alpha)]\mathcal{D}[\hat{a}_\alpha]\hat{\rho}+\kappa_\alpha n^\alpha_{\rm B}(\Omega_\alpha)\mathcal{D}[\hat{a}_\alpha^\dagger]\hat{\rho},
	\end{equation}
	where
	\begin{equation}
	\mathcal{D}[\hat{A}]\hat{\rho}=\hat{A}\hat{\rho}\hat{A}^\dagger-\frac{1}{2}\{\hat{A}^\dagger\hat{A},\hat{\rho}\},
	\end{equation}
	and $\hat{a}_\alpha$ are bosonic annihilation operators and the chemical potentials of both baths are equal to zero.
	
	Calculate the Power and the heat currents in the steady state. Determine when the system acts as a heat engine and when it acts as a refrigerator. Derive the efficiency as well as the coefficient of performance in the respective regimes.
}

\section{Qubit entangler}
\label{sec:entangler}
So far, the examples considered did not exhibit any quantum coherence and can be described by classical rate equations. Here we consider the thermal machine introduced in Ref.~\cite{brask:2015njp}, which uses heat to generate entanglement, a task that simply does not exist in any classical scenario. To this end, we consider two qubits (two-level systems) which are coupled to each other as well as to a thermal bath each with temperatures $T_{\rm C}\leq T_{\rm H}$. The total Hamiltonian then reads
\begin{equation}
\label{eq:hamtotentangler}
\hat{H}_{\rm tot}=\hat{H}_{\rm S}+\sum_{\alpha = \rm C,H}\left(\hat{H}_{\rm \alpha}+\hat{H}_{\rm \alpha S}\right),
\end{equation}
with 
\begin{equation}
\label{eq:hamtotentangler2}
\begin{aligned}
&\hat{H}_{\rm S}=\frac{\varepsilon_{\rm S}}{2}\left(\hat{\sigma}_{\rm C,z}+\hat{\sigma}_{\rm H,z}\right)+g\left(\hat{\sigma}^\dagger_{\rm C}\hat{\sigma}_{\rm H}+\hat{\sigma}^\dagger_{\rm H}\hat{\sigma}_{\rm C}\right),\hspace{1cm}\hat{H}_{\alpha}=\sum_q\varepsilon_{\rm q}\hat{a}_{\alpha,q}^\dagger\hat{a}_{\alpha,q},\\&\hat{H}_{\rm \alpha S}=\sum_q\left(g_{\alpha,q}\hat{\sigma}_\alpha\hat{a}_{\alpha,q}^\dagger+g_{\alpha,q}^*\hat{\sigma}^\dagger_\alpha \hat{a}_q\right),
\end{aligned}
\end{equation}
where we labeled the qubits with a subscript corresponding to the bath they couple to, $\hat{\sigma}_{\alpha,\rm z}$ is the Pauli z-matrix acting on qubit $\alpha$, $\hat{a}_{\alpha,q}$ is a bosonic annihilation operator, and the raising and lowering operators for the qubits read
\begin{equation}
\hat{\sigma}_\alpha^\dagger = |1\rangle_\alpha\langle 0|,\hspace{1.5cm}\hat{\sigma}=|0\rangle_\alpha\langle 1|.
\end{equation}

For this system, the Born-Markov approximations are no longer sufficient to obtain a master equation in GKLS form. A second Markov approximation, as outlined in Sec.~\ref{sec:addapp}, together with the Born-Markov approximations results in the master equation (neglecting the renormalization of $\hat{H}_{\rm S}$)
\begin{equation}
\label{eq:entanglermaster}
\partial_t\hat{\rho}_{\rm S} = -i[\hat{H}_{\rm S},\hat{\rho}_{\rm S}]+\sum_{\alpha=\rm C,H}\mathcal{L}_{\alpha}\hat{\rho}_{\rm S},
\end{equation}
where 
\begin{equation}
\label{eq:bathlindentangler}
\mathcal{L}_\alpha\hat{\rho}=\kappa'_\alpha[1+n_{\rm B}^\alpha(\varepsilon_{\rm S})]\mathcal{D}[\hat{\sigma}_\alpha]\hat{\rho}+\kappa'_\alpha n^\alpha_{\rm B}(\varepsilon_{\rm S})\mathcal{D}[\hat{\sigma}_\alpha^\dagger]\hat{\rho},
\end{equation}
$\kappa'_\alpha = 2\pi\rho_\alpha(\varepsilon_{\rm S})$, and the Bose-Einstein distribution is given at $\mu=0$. Note the local structure of this master equation. The dissipator of each bath only acts on the qubit it is coupled to.

Introducing the quantity
\begin{equation}
\label{eq:kappakappaprime}
\kappa_\alpha = \kappa'_\alpha\frac{n_{\rm B}^\alpha(\varepsilon_{\rm S})}{n_{\rm F}^\alpha(\varepsilon_{\rm S})},
\end{equation}
we can write
\begin{equation}
\label{eq:bathlindentanglerferm}
\mathcal{L}_\alpha\hat{\rho}=\kappa_\alpha[1-n_{\rm F}^\alpha(\varepsilon_{\rm S})]\mathcal{D}[\hat{\sigma}_\alpha]\hat{\rho}+\kappa_\alpha n^\alpha_{\rm F}(\varepsilon_{\rm S})\mathcal{D}[\hat{\sigma}_\alpha^\dagger]\hat{\rho},
\end{equation}
where the Fermi-Dirac distributions are taken at $\mu=0$. For the following discussion, it is important to keep in mind that $\kappa_\alpha$ are temperature-dependent quantities.

The first Markov approximation is justified if
\begin{equation}
\label{eq:markov1entangler}
\rho_{\alpha}(\varepsilon_{\rm S}\pm\gamma)\simeq\rho_{\alpha}(\varepsilon_{\rm S}),\hspace{1cm}n_{\rm B}^{\alpha}(\varepsilon_{\rm S}\pm\gamma)\simeq n_{\rm B}^{\alpha}(\varepsilon_{\rm S}),
\end{equation}
where $\gamma=\kappa_{\rm C}+\kappa_{\rm H}$. The second Markov approximation is justified if
\begin{equation}
\label{eq:markov2entangler}
\rho_{\alpha}(\varepsilon_{\rm S}\pm g)\simeq\rho_{\alpha}(\varepsilon_{\rm S}),\hspace{1cm}n_{\rm B}^{\alpha}(\varepsilon_{\rm S}\pm g)\simeq n_{\rm B}^{\alpha}(\varepsilon_{\rm S}).
\end{equation}
These conditions are usually fulfilled if
\begin{equation}
\label{eq:kapgsmale}
\kappa_{\rm C},\,\, \kappa_{\rm H},\,\, g\ll\varepsilon_{\rm S}.
\end{equation}
We note that it is $\kappa$, not $\kappa'$ which enters these equations. The reason for this is that the qubit, as a two-level system, is described by Fermi-Dirac statistics (since $\{\hat{\sigma},\hat{\sigma}^\dagger\}=1$), not by Bose-Einstein statistics. The time-scale over which the system changes is thus determined by $\kappa$, not $\kappa'$.

\subsection{Steady state}
The steady state solution to Eq.~\eqref{eq:entanglermaster} reads
\begin{equation}
\label{eq:entanglerss}
\begin{aligned}
\hat{\rho}_{\rm S}=&\frac{\kappa_{\rm C}\kappa_{\rm H}}{\kappa_{\rm C}\kappa_{\rm H}+4g^2}\hat{\tau}^{\rm C}_{\beta_{\rm C}}\hat{\tau}^{\rm H}_{\beta_{\rm H}}+\frac{4g^2}{\kappa_{\rm C}\kappa_{\rm H}+4g^2}\hat{\tau}^{\rm C}_{\bar{\beta}}\hat{\tau}^{\rm H}_{\bar{\beta}}\\&+\frac{2g\kappa_{\rm C}\kappa_{\rm H}[n_{\rm F}^{\rm H}(\varepsilon_{\rm S})-n_{\rm F}^{\rm C}(\varepsilon_{\rm S})]}{(\kappa_{\rm C}+\kappa_{\rm H})(\kappa_{\rm C}\kappa_{\rm H}+4g^2)}i\left(\hat{\sigma}_{\rm C}^\dagger\hat{\sigma}_{\rm H}-\hat{\sigma}_{\rm H}^\dagger\hat{\sigma}_{\rm C}\right),
\end{aligned}
\end{equation}
where we introduced the single-qubit thermal states
\begin{equation}
\label{eq:singtherm}
\hat{\tau}_\beta^\alpha = \frac{e^{-\beta\varepsilon_{\rm S}\hat{\sigma}_{\alpha,\rm z}/2}}{{\rm Tr}\{e^{-\beta\varepsilon_{\rm S}\hat{\sigma}_{\alpha,\rm z}/2}\}}=n_{\rm F}^{\beta}(\varepsilon_{\rm S})|1\rangle_\alpha\langle 1|+[1-n_{\rm F}^{\beta}(\varepsilon_{\rm S})]|0\rangle_\alpha\langle 0|,
\end{equation}
with $n_{\rm F}^{\beta}(\varepsilon_{\rm S})$ denoting the Fermi function with inverse temperature $\beta$ and chemical potential $\mu=0$. Furthermore, the temperature $\bar{\beta}$ is defined such that
\begin{equation}
\hat{\tau}^\alpha_{\bar{\beta}}=\bar{n}|1\rangle_\alpha\langle 1|+(1-\bar{n})|0\rangle_\alpha\langle 0|,\hspace{1cm}\bar{n}=\frac{\kappa_{\rm C}n_{\rm F}^{\rm C}(\varepsilon_{\rm S})+\kappa_{\rm H}n_{\rm F}^{\rm H}(\varepsilon_{\rm S})}{\kappa_{\rm C}+\kappa_{\rm H}}.
\end{equation}
The first line in Eq.~\eqref{eq:entanglerss} is purely diagonal in the computational basis $|n_{\rm C},n_{\rm H} \rangle$ where $n_\alpha=0,1$. The second line however represents coherence between the two qubits since the term is proportional to
$|1_{\rm C}0_{\rm H}\rangle\langle 0_{\rm C}1_{\rm H}|-|0_{\rm C}1_{\rm H}\rangle\langle 1_{\rm C}0_{\rm H}|$. This term is finite as long as $n_{\rm F}^{\rm H}(\varepsilon_{\rm S})\neq n_{\rm F}^{\rm C}(\varepsilon_{\rm S})$ and thus requires an out-of-equilibrium situation. In this case, a heat flow through the system induces coherences between the qubits. As discussed below, these coherences can become strong enough that the two qubits become entangled.

It is illustrative to consider some limiting cases. First, we consider the case of a vanishing interaction
\begin{equation}
g\rightarrow 0\hspace{.5cm}\Rightarrow\hspace{.5cm}\hat{\rho}_{\rm S}=\hat{\tau}^{\rm C}_{\beta_{\rm C}}\hat{\tau}^{\rm H}_{\beta_{\rm H}}.
\end{equation}
 This state describes two uncoupled qubits, each having thermalized with its respective thermal bath. The second limiting case we consider is the case of thermal equilibrium (remember that $\mu=0$ here)
 \begin{equation}
 \label{eq:nottaub}
T_{\rm C}=T_{\rm H}\hspace{.5cm}\Rightarrow\hspace{.5cm}\hat{\rho}_{\rm S}=\hat{\tau}^{\rm C}_{\beta}\hat{\tau}^{\rm H}_{\beta}.
 \end{equation}
 This state is the same as the one obtained from two uncoupled qubits each thermalizing with a bath of the same temperature. We note that this state is different from the thermal state $\hat{\tau}_{\beta}$ because the latter includes the coupling term between the two qubits. This is a consequence of the local structure of the master equation. We note that under the conditions in Eq.~\eqref{eq:markov2entangler}, which ensure that the master equation is justified, the differences between $\hat{\tau}_{\beta}$ and the state in Eq.~\eqref{eq:nottaub} are vanishingly small.
 The final limiting case we consider is the case where the internal coupling is much larger than the coupling to the baths
 \begin{equation}
 \kappa_{\rm C}/g,\kappa_{\rm H}/g\rightarrow 0\hspace{.5cm}\Rightarrow\hspace{.5cm}\hat{\rho}_{\rm S}=\hat{\tau}^{\rm C}_{\bar{\beta}}\hat{\tau}^{\rm H}_{\bar{\beta}}.
 \end{equation}
 This state again looks like the state obtained in thermal equilibrium with inverse bath temperatures determined by $\bar{\beta}$. This is analogous to the quantum dot example discussed in Sec.~\ref{sec:qdhe}. Also there did we find that a system coupled to two baths with different temperatures behaves just like the same system coupled to a single bath with mean occupation given by $\bar{n}$, when only considering the state of the system. Of course, there is a big difference between in- and out-of-equilibrium situations when considering energy flows.
 
\subsection{Entanglement}
\begin{figure}
	\centering
	\includegraphics[width=.6\textwidth]{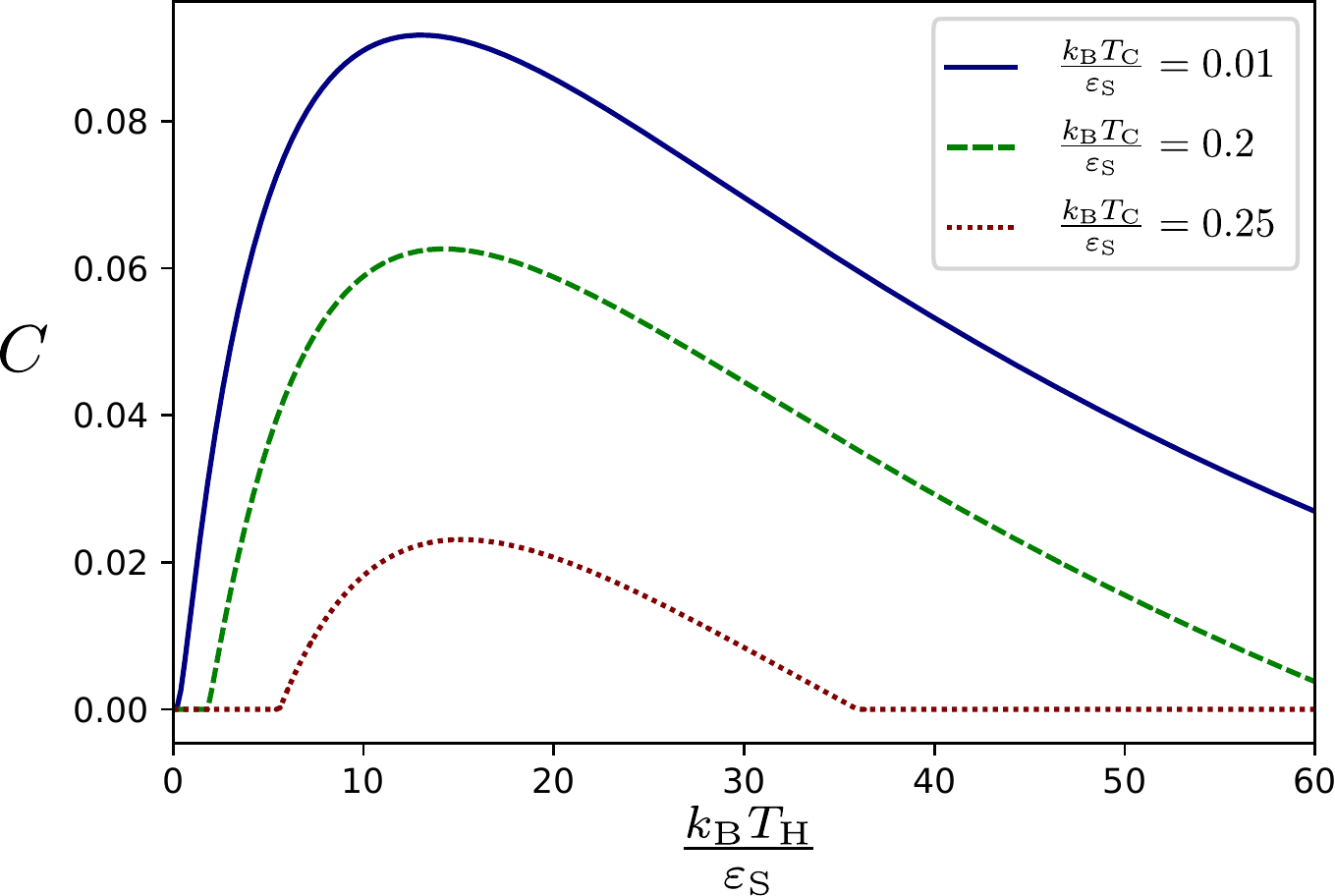}
	\caption{Concurrence as a function of the hot temperature for different values of the cold temperature. For $T_{\rm C}\rightarrow0$, we find a finite concurrence for small but finite $T_{\rm H}$. The concurrence then goes through a maximum and diminishes at high $T_{\rm H}$. This reduction of the concurrence is due to the fact that $\kappa_{\rm H}$ grows linearly in $T_{\rm H}$ for large hot temperatures. For finite $T_{\rm C}$, we find a threshold $T_{\rm H}$ below which there is no entanglement, in analogy to Ref.~\cite{brask:2015njp}. Parameters: $g/\varepsilon_{\rm S}=0.02$, $\kappa'_{\rm C}/\varepsilon_{\rm S}=0.1$, $\kappa'_{\rm H}/\varepsilon_{\rm S}=0.001$.}
	\label{fig:entangler}
\end{figure}

\begin{tcolorbox}
	\paragraph{Entanglement, entanglement of formation, and concurrence} Entanglement is arguably the most intriguing feature of quantum mechanics. To define entanglement, a partition is required such that a part of the state can be attributed to Alice (A) and a part to Bob (B). A state is then said to be entangled if it cannot be written in the form
	\begin{equation}
	\label{eq:separable}
	\hat{\rho} = \sum_j p_j\hat{\rho}_{\rm A}^j\otimes\hat{\rho}_{\rm B}^j, \hspace{2cm}\sum_jp_j=1,
	\end{equation}
	where $p_j\geq0$ and $\hat{\rho}_{\rm A/B}^j$ are density matrices in the subspace of Alice/Bob. Any state which \textit{can} be written in this form is called separable. 
	
	The amount of entanglement can be quantified by the \textit{entanglement of formation} \cite{bennett:1996}. Loosely speaking, it is determined by the number of Bell states, i.e., states of the form $\tfrac{1}{\sqrt{2}}(|01\rangle+|10\rangle)$, that are required to prepare the given state using only local operations and classical communication (LOCC). For two qubits, the entanglement of formation is a number between zero and one, where zero is obtained for separable states and one for Bell states.
	
	Determining if a given state is entangled is in general a highly non-trivial task. However, for two qubits the low dimensionality of the problem considerably facilitates the task. A common measure for entanglement in this scenario is the \textit{concurrence} \cite{hill:1997,wootters:1998}, which is monotonically related to the entanglement of formation. Just as the latter, the concurrence ranges from zero, obtained for separable states, to one, reached for Bell states.
	
\end{tcolorbox}
As can be seen from Eq.~\eqref{eq:entanglerss}, a non-equilibrium situation can give rise to coherences between the two qubits. If these are strong enough, the state becomes entangled. The amount of entanglement can be characterized by the concurrence (see box), which can be calculated analytically and reads
\begin{equation}
\label{eq:concurrencerhs}
\begin{aligned}
C[\hat{\rho}_{\rm S}]=&\frac{2\kappa_{\rm C}\kappa_{\rm H}}{4g^2+\kappa_{\rm C}\kappa_{\rm H}}\left\{\frac{2g}{\kappa_{\rm C}+\kappa_{\rm H}}\left(n_{\rm F}^{\rm H}-n_{\rm F}^{\rm C}\right)\right.\\&\left.-\sqrt{\left[(1-n_{\rm F}^{\rm C})\left(1-n_{\rm F}^{\rm H}\right)+\frac{4g^2}{\kappa_{\rm C}\kappa_{\rm H}}\left(1-\bar{n}\right)^2\right]\left[n_{\rm F}^{\rm C}n_{\rm F}^{\rm H}+\frac{4g^2}{\kappa_{\rm C}\kappa_{\rm H}}\bar{n}^2\right]}\right\},
\end{aligned}
\end{equation}
if this quantity is positive. Otherwise, the concurrence is equal to zero. Here we suppressed the arguments of the Fermi-Dirac distributions for ease of notation. We note that $\kappa_\alpha$ are temperature dependent parameters. The concurrence is plotted in Fig.~\ref{fig:entangler} as a function of the hot temperature, for different values of the cold temperature.

The fact that coupling a system to thermal baths can give rise to entanglement is remarkable  due to the following reason: when coupling to a bath, the qubits in the system get entangled with the degrees of freedom in the bath. However, entanglement is \textit{monogamous}, which means that if a qubit is strongly entangled with some degree of freedom, it cannot at the same time be strongly entangled with another degree of freedom \cite{coffman:2000}. This is why the presence of an environment usually suppresses inter-system entanglement. The qubits in the system have to \textit{share} their entanglement with all the bath modes, leaving little entanglement for themselves. Here we encounter a different behavior, where the bath induces entanglement between the two system qubits.

{\small 
	\paragraph{Exercises}
	
	\paragraph{3.2 Markovian master equations: comparing approximations (4pt)}
	Consider the qubit entangler described by Eqs.~\eqref{eq:hamtotentangler} and \eqref{eq:hamtotentangler2}, where the chemical potential of both reservoirs is equal to zero. Derive Markovian master equations under Born-Markov approximations using the secular approximation, the strong Markov approximation, as well as the PERLind approach discussed in Sec.~\ref{sec:addapp}. Compute and plot the heat currents as a function of the inter-system coupling strength $g$, neglecting the renormalization of the Hamiltonian and assuming energy-independent bath spectral densities. Compare the three approaches and comment on their regime of validity.
}

\section{Conclusions \& outlook}
In these lecture notes, we gave a brief introduction into the rapidly growing field of quantum thermodynamics. In Sec. \ref{sec:laws}, we discussed how the laws of thermodynamics arise in quantum theory and discussed how the concept of entropy is intimately related to a the lack of knowledge of microscopic degrees of freedom. Loosely speaking, the second law then states that our knowledge is decreasing over time. While this is true for large and memoryless baths, it breaks down when the baths are small or when non-Markovian effects become important. The investigation of such scenarios, which are not part of these lecture notes, is an active field of research \cite{strasberg:2019}.

The rest of the lecture notes focused on Markovian master equations. We discussed how such a description is obtained from microscopic equations of motions, using Born-Markov as well as additional approximations. While Markovian master equations provide a powerful tool and are ubiquitous in describing open quantum systems, it is important to keep their limitations in mind as they can give erroneous results, such as a violation of the second law, when employed inappropriately. While Markovian master equations have been widely studied for many decades, the current interest in open quantum systems out of equilibrium led to a renewed interest in this fundamental topic because some conclusions drawn for equilibrium reservoirs are no longer true when temperature and/or voltage biases are involved \cite{hofer:2017njp,mitchison:2018,gonzalez:2017,seah:2018,kirsanskas:2018}.

We illustrated the use of Markovian master equations at the example of a quantum dot heat engine, as well as a qubit entangler. These are only two examples out of a variety of thermal machines that have been considered. While the most prominent tasks for thermal machines are the production of heat or refrigeration, other task, such as measuring temperature \cite{hofer:2017}, or keeping time \cite{erker:2017} have been considered.

Quantum thermodynamics is a quickly growing field, with new results appearing on the ArXiv on a weekly if not daily basis. Many topics were thus not considered in these notes. Here we briefly comment on a few of those topics to illustrate the broad scope of the field:
\begin{enumerate}
	\item \textbf{Fluctuations}: Throughout these notes, we only considered the mean value of heat currents and power. However, these currents will in general be fluctuating quantities that vary in time. In classical scenarios, the investigation of fluctuations has resulted in a number of powerful results such as fluctuation theorems \cite{seifert:2012} and thermodynamic uncertainty relations \cite{barato:2015,gingrich:2016}. While generalizations to the quantum regime exist, this regime is much more involved because of the unavoidable measurement backaction. In quantum mechanics, it is in general impossible to observe a fluctuating quantity over time without influencing its behavior. Indeed, this quantum effect resulted in ongoing debates on the definition of work as a fluctuating quantity \cite{thermo:book}.
	\item \textbf{Thermodynamic cycles}: The focus of these notes is on thermal machines that operate in the steady state. Such machines are often called \textit{autonomous} machines, at least if they only require static voltage and temperature biases for their operation. Different types of thermal machines are provided by quantum versions of thermodynamic cycles such as the Otto or the Carnot cycle. While such a cyclic operation usually requires a higher degree of control, the investigation of quantum versions of thermodynamic cycles is a promising route towards understanding the subtle differences between stochastic (i.e., classical but fluctuating) and quantum thermodynamics \cite{thermo:book}.
	\item \textbf{Thermodynamics of information}: This is a subject that goes back to the thought experiments of Maxwell \cite{maxwell:1871} and Szilard \cite{szilard:1929}. Considering the close relation between entropy and information, it is not surprising that measurement and feedback protocols can strongly affect thermodynamic properties. A proper thermodynamic bookkeeping of information is thus indispensable to describe such scenarios \cite{parrondo:2015,goold:2016}. Recent developments in this rapidly growing subject include an investigation on quantum many-body effects in a Szilard engine \cite{bengtsson:2018}, experiments on Maxwell demon scenarios in the quantum regime \cite{cottet:2017,masuyama:2018,naghiloo:2018}, and the extension of fluctuation theorems \cite{sagawa:2010,potts:2018} as well as thermodynamic uncertainty relations \cite{potts:2019} to include measurement and feedback scenarios. 
	\item \textbf{Resource theory of quantum thermodynamics}: This is a theoretical framework built on a resource theory, similar in spirit to the resource theory of entanglement \cite{lostaglio:2018rev,chitambar:2019}. It is a fairly abstract approach which is very different in nature from the approach taken in these lecture notes. The resource theory of quantum thermodynamics has resulted in a number of insights into what is fundamentally possible using thermal reservoirs. While these results can seem fairly abstract, recent works have started to connect them to experimental scenarios \cite{yunger:2017,lorch:2018,lostaglio:2018,yunger:2019}. 
\end{enumerate}

\paragraph{Acknowledgments} I thank all participants of the Quantum Thermodynamics course at Lund University in the Spring Semester 2019. The lectures were often followed by inspiring discussions, in particular with Peter Samuelsson, Claudio Verdozzi, and Marcus Dahlstr\"om. I acknowledge funding from the European Union's Horizon 2020 research and innovation programme under the Marie Sk{\l}odowska-Curie Grant Agreement No. 796700.

%\section{Time-dependent Hamiltonians}
%\begin{itemize}
%	\item Master equations
%	\item Heat and Work
%	\item Quantum Heat engine \cite{hofer:2016prb}
%\end{itemize}
%
%\section{Fluctuations$^*$}
%\begin{itemize}
%	\item Full-counting statistics
%	\item Fluctuation theorems (?)
%\end{itemize}
%
%\section{The Resource Theory of Quantum Thermodynamics$^*$}
%\begin{itemize}
%	\item Work extraction
%	\item From resource theory to experiment
%\end{itemize}
%
%
%
%\section*{What is missing}
%\begin{itemize}
%	\item Thermodynamic cycles
%	\item Equilibration
%	\item Quantum Trajectory approach
%	\item Scattering approach
%	\item Strong coupling thermodynamics
%	\item Thermometry
%	\item Thermodynamics of information
%	\item Thermodynamics of measurements
%\end{itemize}

\bibliographystyle{quantum_ph}
\bibliography{biblio}
\end{document}